\documentclass[12pt]{article}
\pdfoutput=1

\usepackage[
 linktoc=all,
 colorlinks=true,
 linkcolor=blue,
 urlcolor=blue,
 citecolor=red
 ]{hyperref}

\usepackage[backend=bibtex,style=numeric-comp,sorting=none,maxbibnames=99, minbibnames=99]{biblatex}
\usepackage{mathtools,amssymb,mathrsfs,tikz,dsfont}
\usetikzlibrary{decorations.pathmorphing}
\usetikzlibrary{decorations.markings}

\usepackage{microtype}

\DeclareMathOperator{\diag}{diag}

\def\be{\begin{eqnarray}}
\def\ee{\end{eqnarray}}
\newcommand{\nn}{\nonumber}
\newcommand\para{\paragraph{}}
\newcommand{\ft}[2]{{\textstyle\frac{#1}{#2}}}
\newcommand{\eqn}[1]{\eqref{#1}}

\newcommand{\ra}{\rightarrow}

\def\Dslash{\,\,{\raise.15ex\hbox{/}\mkern-12mu D}}
\def\Dbarslash{\,\,{\raise.15ex\hbox{/}\mkern-12mu {\bar D}}}
\def\delslash{\,\,{\raise.15ex\hbox{/}\mkern-9mu \partial}}
\def\delbarslash{\,\,{\raise.15ex\hbox{/}\mkern-9mu {\bar\partial}}}
\def\pslash{\,\,{\raise.15ex\hbox{/}\mkern-9mu p}}
\def\calDslash{\,\,{\raise.15ex\hbox{/}\mkern-12mu {\cal D}}}

\newcommand{\1}{\mathbf{1}}

\newcommand{\Z}{{\mathds Z}}

\newcommand{\R}{{\mathds R}}

\newcommand{\ket}{\rangle}

\newcommand{\bx}{{\bf x}}

\newcommand{\kett}[1]{|#1\rangle}

\def\lae{\mathrel{\mathop{\smash{\lower .5 ex \hbox{$\stackrel<\sim$}}}}}
\def\lae{\mathrel{\mathop{\smash{\lower .5 ex \hbox{$\stackrel>\sim$}}}}}

\usepackage[framemethod=TikZ]{mdframed}
\mdfsetup{%
skipabove=3pt,
skipbelow=2pt,
linecolor=black!0,
backgroundcolor=black!15,
roundcorner=3pt}

\renewcommand{\title}[1]{\vbox{\center\LARGE{#1}}\vspace{5mm}}
\renewcommand{\author}[1]{\vbox{\center\large#1}\vspace{5mm}}
\newcommand{\address}[1]{\vbox{\center\em#1}}

\makeatletter\@addtoreset{equation}{section}\makeatother

\usepackage{moresize}

\begin{document}

\begin{titlepage}

\begin{center}
\vspace{5mm}
\hfill {\tt }\\
\vspace{8mm}
\title{\makebox[\textwidth]{\fontsize{25}{40}\selectfont Fermion-Monopole Scattering}}
\vspace{-7mm}
\title{\makebox[\textwidth]{\fontsize{25}{40}\selectfont in the Standard Model}}

\vspace{10mm}
 Marieke van Beest,${}^a\,$\footnote{\href{mailto:mvanbeest@scgp.stonybrook.edu}{\tt mvanbeest@scgp.stonybrook.edu}}
 Philip Boyle Smith,${}^b\,$\footnote{\href{mailto:philip.boyle.smith@ipmu.jp}{\tt philip.boyle.smith@ipmu.jp}}
 Diego Delmastro,${}^a\,$\footnote{\href{mailto:ddelmastro@scgp.stonybrook.edu}{\tt ddelmastro@scgp.stonybrook.edu}}\\
 Rishi Mouland,${}^c\,$\footnote{\href{mailto:r.mouland@damtp.cam.ac.uk}{\tt r.mouland@damtp.cam.ac.uk}}
 David Tong.${}^c\,$\footnote{\href{mailto:d.tong@damtp.cam.ac.uk}{\tt d.tong@damtp.cam.ac.uk}}
 \vskip 7mm
 \address{
 ${}^a$ Simons Center for Geometry and Physics,\\
 SUNY, Stony Brook, NY 11794, USA}
 \address{
 ${}^b$ Kavli Institute for the Physics and Mathematics of the Universe (WPI),\\ University of Tokyo, Kashiwa, Chiba 277-8583, Japan}
 \address{
 ${}^c$ Department of Applied Mathematics and Theoretical Physics,\\ University of Cambridge, CB3 0WA, UK}
\end{center}

\vspace{5mm}

\abstract{
\noindent We study the scattering of fermions off 't Hooft lines in the Standard Model. A long-standing paradox suggests that the outgoing fermions necessarily carry fractional quantum numbers. In a previous paper, we resolved this paradox in the context of a number of toy models  where we showed that the outgoing radiation is created by operators that are attached to a co-dimension 1 topological surface. This shifts the quantum numbers of the outgoing states associated to non-anomalous symmetries to be integer valued as required, while the quantum numbers associated to anomalous symmetries are fractional. Here we apply these ideas to the Standard Model. 
}
 
\vfill\eject


\end{titlepage}

{\hypersetup{linkcolor=black}
\tableofcontents
\thispagestyle{empty}
}

\unitlength = .8mm

\setcounter{tocdepth}{3}

\section{Introduction}

Monopoles and chiral gauge theories make for a heady, and poorly understood, mix. The fun happens when you throw a chiral fermion at a monopole. Under certain circumstances, it naively appears that there is nothing sensible that can bounce back~\cite{doi:10.1063/1.2947547,VARubakov_1988,Callan:1982ah,Rubakov:1982fp,PhysRevD.28.876,PhysRevLett.52.1755,doi:10.1063/1.34591,doi:10.1146/annurev.ns.34.120184.002333,Polchinski:1984uw}.

\para
More precisely, this puzzle arises when one focuses on the lowest angular momentum sector of the scattering. The lowest angular momentum mode is special for two reasons. First, this mode experiences no angular momentum barrier and so penetrates into  the core of the monopole where it is at the whim of whatever UV physics awaits. Second, each Weyl fermion contributes \emph{either} an ingoing mode \emph{or} an outgoing mode in this sector. This is in contrast to the higher angular momentum sectors where each fermion gives rise to both ingoing and outgoing degrees of freedom. In a chiral gauge theory, where left-handed and right-handed fermions experience different forces, we therefore find ourselves in a situation where ingoing scattering modes carry different quantum numbers from the outgoing modes. The question of fermion scattering becomes: how can we patch these modes together, preserving the relevant symmetries of the problem? 

\para
At first glance, the  answer seems to be: we can't. This was first pointed out by Callan in the context of the Standard Model \cite{doi:10.1063/1.34591}. He showed that if you throw a right-handed electron $e_R$ at a point-like Dirac monopole, then the only thing that can bounce back preserving gauge quantum numbers is
\be e_R \longrightarrow \frac{1}{2}\left(\bar{u}_R^1 + \bar{u}_R^2 + \bar{q}_L^3 + l_L\right)\ .\label{puzzle}\ee
Here $l_L$ and $q_L$ are left-handed leptons and quarks respectively, both in the doublet of $SU(2)_\text{weak}$, and $u_R$ is a right-handed up quark. The superscripts 1, 2, and 3 are colour indices, arranged so that the outgoing state is a colour singlet.

\para
The mystery lies in that overall factor of $1/2$ in \eqn{puzzle}. What to make of it? It's not  possible for, say, half an up  quark to bounce back. So what does it mean? The purpose of this paper is to resolve this puzzle.

\para
The resolution  is surprising.  Said in a heuristic, and slightly dramatic, way, the monopole provides a portal into a different, twisted, Hilbert space, related to the original by the action of a discrete symmetry. More concretely, the outgoing state is created by an operator that is  attached to a topological surface  which connects it to the monopole. This proposal was made recently in  \cite{vanBeest:2023dbu}. Related ideas were previously put forward in \cite{Hamada:2022eiv}  and, in the 2d context, in \cite{Maldacena:1995pq}. Other proposals for the resolution the puzzle can be found in \cite{Affleck:1993np,Brennan:2021ewu,Csaki:2021ozp,Csaki:2022qtz,Csaki:2022tvb,Kitano:2021pwt,Khoze:2023kiu,Brennan:2023tae}.

\para
The  fermion-monopole problem was resolved  in \cite{vanBeest:2023dbu} for a  simple model of gapless quantum electrodynamics and not for the Standard Model itself. In this paper, we first delineate the fermion scattering problem for all possible monopoles in the Standard Model. These are point-like Dirac monopoles, also known as 't Hooft lines, that we put in by hand, rather than solitonic 't Hooft Polyakov monopoles~\cite{Polyakov:1974ek,tHooft:1974kcl}. By ``all possible'' monopoles, we mean those carrying different magnetic fluxes in all factors of the $SU(3)\times SU(2) \times U(1)$ gauge group. 

\para
Optimistically, one may hope that the prescription of attaching a topological surface defect to the outgoing fermions will resolve the fermion-monopole scattering paradox for any chiral gauge theory. In practice, the number of theories for which we can uniquely determine these surface operators is rather limited and so, for a  general chiral gauge theory, the problem remains open.  Moreover, for a general chiral theory it appears that the surface defect must somehow be associated to the $SU(2)_\text{rot}$ rotational symmetry in a way that remains mysterious (we will elaborate on this point in section~\ref{whatsec}). All of which prompts the question: what happens in the Standard Model?

\para
For most choices of magnetic charges in $SU(3)\times SU(2)\times U(1)$, we are unable to construct a suitable surface defect and we cannot answer the question of what bounces back.  There are, however, exceptions and, happily, this arises for the case that one would naively call \emph{the} magnetic monopole, meaning the one that carries minimal magnetic charge after electroweak symmetry breaking. Here there is a fortuitous  accident that allows us to construct the appropriate surface operator. We will see that the  seemingly nonsensical scattering process \eqn{puzzle} instead becomes
\be e_R \longrightarrow l_L  + T\label{resolved}\ee
%
%
%
Here  $T$ denotes a twist operator for a particular  $\Z_2$ symmetry. It carries various charges under the symmetries of the problem that we will enumerate. These ensure that  conservation laws for all non-anomalous symmetries are obeyed. This includes both the gauge quantum numbers and $B-L$. However,  baryon and lepton number which, individually, are anomalous are necessarily violated in this scattering. This manifests itself by the twist operator carrying fractional $B$ and $L$ charge. This violation of baryon number is what underlies proton decay in the famous Callan-Rubakov effect \cite{Rubakov:1982fp,Callan:1982au,Callan:1982ac,VARubakov_1988}. 

\para
One novelty vis-\`a-vis~\cite{vanBeest:2023dbu} that arises in the Standard Model is that there is a multiplet of twist operators, each carrying different charges. This arises from the presence of fermionic zero modes. Curiously, one of these twist operators is found to carry the same quantum numbers as the Higgs expectation value, and this is what ultimately allows the scattering process \eqn{resolved} to fly. 



\para
The paper is organised as follows. In section~\ref{knowsec}, we set the scene. We describe the basic phenomenon of Weyl fermions scattering off a monopole and explain what is special about the lowest angular momentum sector. We give a number of simple examples that highlight the kind of scattering problems that we can solve, and those where we do not  (yet) have the technology to construct the relevant boundary conditions and surface operators.

\para
In section~\ref{monosec}, we look at the various 't Hooft lines in the Standard Model. For each, we determine the quantum numbers of the ingoing and outgoing fermions. In other words, in this section we pose the problem that we would like to solve. In section~\ref{nicesec}, we focus on \emph{the} magnetic monopole, which we will refer to as the ``minimal monopole'',  where we can provide an answer. We will construct the multiplet of twist operators and show that, for each ingoing state, there is an appropriate outgoing state. Finally, in chapter~\ref{whatsec}, we provide some thoughts on what it means for the outgoing radiation to be attached to a defect operator and highlight some open questions.

\section{Fermion-Monopole Scattering}\label{knowsec}

In this section we recount some basic facts about the scattering of fermions off monopoles and review the results of \cite{vanBeest:2023dbu}. We also explain the distinction between those theories where we can construct surface operators and those where, at least for now, we cannot.

\para
The key piece of physics underlying the whole story is a cute observation from classical mechanics. Take a particle of electric charge $q$ in the presence of a monopole of magnetic charge $g$. Then the conserved angular momentum ${\bf L}$ picks up an additional, anomalous term \cite{poincare}
\be {\bf L} = \bx\times{\bf p}- \frac{qg}{2}\hat{\bf r}\ .\label{poincare}\ee
This same shift of the angular momentum arises in field theory. Consider a single left-handed Weyl fermion $\psi$. The free fermion is invariant under a global $U(1)$ symmetry. For future applications, it's useful to take the fermion to have charge $q\in \Z$, so 
\be \psi \ra e^{iq\alpha} \psi\ .\ee
We couple this symmetry to a background gauge field $A$ in which we turn on the configuration of a spherically symmetric magnetic monopole,
\be A = \frac{g}{2} (1-\cos\theta) d\phi \ \ \ \mbox{with $g\in \Z$}\ .\ee
We then decompose the Weyl spinor $\psi$ into monopole harmonics \cite{WU1976365,PhysRevD.15.2287}.  (A detailed summary of this calculation can be found in \cite{Hamada:2022eiv}.) This is where things get interesting. The lowest lying angular momentum mode has total angular momentum
\be j_0 = \frac{|gq|-1}{2}\ .\ee
Here the $-1/2$ is just the spin of the fermion, while the factor of  $gq/2$ can be traced to the anomalous term in  \eqn{poincare}. This means, among other things, that a charge $q=1$ fermion moving in the presence of a charge $g=1$ monopole has its lowest mode as the s-wave with $j_0=0$. Indeed, whenever $gq$ is odd, the angular momentum modes  of the fermion are all integer, rather than half-integer.

\para
Crucially, the lowest angular momentum mode is chiral: the mode is:
\begin{itemize}
\item Ingoing for $qg>0$.
\item Outgoing for $qg<0$.
\end{itemize}
This is the fact that lies at the heart of the scattering paradox and it is true only for the lowest angular momentum mode: all higher modes come in ingoing/outgoing pairs. This is entirely analogous to the statement that the lowest Landau level of graphene is spin polarised, while all higher Landau levels are not. 

\para
In addition, each of the higher modes experiences an angular momentum barrier. If we denote the $j^\text{th}$ angular momentum mode as $\psi_j$ then, near the core of the monopole at $r=0$, the profile takes the form
\be  \psi_j(r) \sim r^{\nu-1} \ \ \ \text{with}\ \ \ \nu = \sqrt{(j-j_0)(j+j_0+1)}\ .\ee
With $\psi_j$ viewed as a wavefunction, we can interpret $r^2 |\psi_j|^2 \sim r^{2\nu}$ as the probability that the $j^\text{th}$ angular momentum mode makes it into the core of the monopole. We see that only the lowest angular momentum mode with $j=j_0$ and, correspondingly,  $\nu=0$, has non-vanishing probability of making it to the middle. 

\para
This makes sense. The higher angular momentum modes come in ingoing/outgoing pairs and, at low energy, simply bounce off the angular momentum barrier with a phase shift. It is only the lowest angular momentum modes with $j=j_0$ that experience the heart of the monopole. (One might worry that this power-law decay isn't strictly sufficient for us to restrict to the lowest-angular momentum sector, which might require an exponential fall-off. At weak coupling, the restriction to the lowest-angular momentum does appear to hold. See, however, \cite{Csaki:2021ozp}, for a different proposal.) 

\para
With this background in mind, we can now look at various examples of fermion-monopole scattering. These examples fall into three categories: the understood, the not understood, and the impossible. 

\subsection{Examples of Scattering}\label{examplesec}

We start with the impossible. Take a single Weyl fermion $\psi$ with charge $q=1$ in the background of a magnetic monopole of charge $g=1$. The lowest angular momentum mode is an s-wave with $j_0=0$. There is a single ingoing mode, and no outgoing mode. Clearly, if we throw in a fermion in the s-wave, there is nothing that can bounce back.



\para
This is not a paradox; it is merely a fact. The $U(1)$ global symmetry has a 't Hooft anomaly and therefore the current obeys an anomalous conservation law. Specifically, this means that the current is not strictly conserved in the presence of topologically non-trivial backgrounds. In other words, the 't Hooft line explicitly breaks $U(1)$ and hence the monopole absorbs charge (or pumps it out, depending on the chirality of the fermion). The fact that there is no scattering consistent with the $U(1)$ symmetry is a manifestation of this anomaly. Indeed, in this case the monopole must also absorb energy, which is a manifestation of the mixed gauge-gravitational anomaly.

\para
Said slightly differently, the 't Hooft anomaly is also an obstacle to gauging the symmetry: Maxwell theory coupled to a single Weyl fermion is a sick theory. (Not ``sick'' as in ``good''. ``Sick'' as in ``bad''.)  The inconsistency of scattering off a monopole provides a particularly physical way to see this sickness. See also~\cite{chen2023boundary} for related recent analysis of the interplay between anomalies and scattering.

\para
There is a similar story when the gauge anomaly vanishes, but the mixed gauge-gravitational anomaly does not. Consider, for example, the ``taxicab'' theory with four Weyl fermions, all left-handed, with charges $q=1,-9,-10,12$. These charges are chosen so that the $U(1)^3$ 't Hooft anomaly vanishes because, as Ramanujan noted,
\be 1^3 + 12^3 = 9^3 + 10^2 = 1729\ .\ee
In the background of a monopole of charge $g=1$, the incoming and outgoing modes sit in the following representations of the $SU(2)_\text{rot}$ rotational group and $U(1)$ global symmetry
\begin{equation}
\begin{array}{l|cc}
&U(1) &SU(2)_\text{rot}\\\hline
\text{Incoming:}\quad& +1 & {\bf 1} \\ & +12 & {\bf 12} \\
\hline
\text{Outgoing:}\quad& +9 & {\bf 9} \\ & +10 & {\bf 10} \\
\end{array}
\end{equation}
%
%
There are also the conjugate fermions which, because the $SU(2)$ representations are (pseudo)-real, take the form  ${\bf 1}_{-1}$ and ${\bf 12}_{-12}$ for ingoing and ${\bf 9}_{-9}$ and ${\bf 10}_{-10}$ for outgoing. Upon restricting to these lower angular momentum modes, we are left with a problem in 2d: can we find conformal boundary conditions that reflect the ingoing modes to the outgoing modes, preserving $SU(2)_\text{rot} \times U(1)$? The answer is no. This is because the 2d theory has a gravitational anomaly, with $c_L-c_R = -6$ and no  such reflecting boundary condition is possible. It is not possible to put a magnetic monopole in a theory with a mixed gauge-gravitational anomaly while preserving both charge and energy \cite{vanBeest:2023dbu}. It is possible to do this preserving charge, but the line must then absorb energy, or vice-versa.

\para
There is a general story at play here, involving the relation between anomalies and boundary conditions \cite{Han_2017,Jensen:2017eof,Thorngren:2020yht,Hellerman:2021fla,Choi:2023xjw}. The restriction to a given angular momentum mode, such as the s-wave, means that the scattering off the monopole becomes effectively a problem on the half-line, with the monopole acting as a boundary. The presence of a 't Hooft anomaly  means that there can be no conformal boundary that preserves the symmetry.  Hence, any symmetry that is anomalous after 2d reduction will either be broken by the 't Hooft line, or the line itself will not conserve energy. In either case, the monopole will absorb the fermion that is thrown in.

\para
This means that to frame the paradox in which fermions scatter elastically, we need to look at a magnetic monopole in a theory in which anomalies are absent. Here we describe two: one in which we can solve the scattering, and one in which we can't.

\para
First, the situation where we will declare success. Consider a $U(1)$ gauge theory with an $SU(N)$ flavour symmetry, with left-handed Weyl fermions in the following representations
\begin{equation}
\begin{array}{l|cc}
&U(1) &SU(N)\\\hline
\psi\quad& +1 & {\bf N} \\ 
\tilde{\psi} & -1 & {\bf N} \\
\end{array}
\label{sqed}\end{equation}
%
%
This content arises naturally from an $SU(2)$ gauge theory, broken to $U(1)$ by an adjoint Higgs. In that context, $N$ must be even to avoid the Witten anomaly~\cite{WITTEN1982324}.

\para
This is the theory that, historically, has been used to study the fermion-monopole scattering paradox. In the presence of a charge $g=1$  monopole, the lowest angular momentum modes are again in the s-wave, and sit in the following representations of the $U(1)\times SU(N)$ symmetry
\begin{equation}
\begin{array}{l|cc}
&U(1) &SU(N)\\\hline
\text{Incoming:}\quad& +1 & {\bf N} \\  \hline
\text{Outgoing:} & -1 & {\bf N} \\
\end{array}
\label{sun}\end{equation}
%
This makes the puzzle clear: if you send in an s-wave fermion $\psi$ in the ${\bf N}_{+1}$, then you can naively preserve \emph{either} the $U(1)$ symmetry by returning a fermion $\tilde{\psi}^\dagger$ in the $\overline{\bf N}_{+1}$, \emph{or}  you can preserve the $SU(N)$ symmetry by returning a fermion $\tilde{\psi}$ in the ${\bf N}_{-1}$. But you seemingly cannot preserve both.

\para
Given our previous comments, you might think that the $SU(N)$ symmetry is expendable here. After all, it suffers an $SU(N)^3$ 't Hooft anomaly, which precludes the existence of a 3-dimensional boundary preserving the symmetry. However, the monopole corresponds to a 1-dimensional boundary, and the cubic anomaly has no bearing on the existence of such a symmetric boundary state. From the 2d perspective, the group $SU(N)$ is non-anomalous -- the $SU(N)^2$ 't Hooft anomaly vanishes -- and, indeed, it is possible to construct a 2d conformal boundary condition that preserves both $U(1)$ and $SU(N)$ \cite{Affleck:1993np}. The question is: how should we interpret the scattering?

\para
The answer given in \cite{vanBeest:2023dbu} is that the outgoing state manifestly preserves the $SU(N)$ symmetry, but is created by a combination of a local fermionic operator and a twist operator,
\be \psi \longrightarrow \tilde{\psi} + T^{-2}\ .\label{twist}\ee
The twist operator is attached to a topological line $T$ which generates a $\Z_N \subset U(1)\times SU(N)$ symmetry acting on the outgoing fermions: 
\be T: \psi \ra \psi\ \ \ \text{and}\ \ \ \tilde{\psi} \ra \omega \tilde{\psi}\ \ \ \text{with}\ \ \ \omega  = e^{2\pi i/N} .\ee
The reduction to the s-wave allows us to think of scattering in a purely two-dimensional sense. From this perspective, the  meaning of  the twist operator $T^{-2}$ in \eqn{twist} is that the outgoing fermion $\tilde{\psi}$ is connected to the monopole boundary by a topological line as shown in the figure below:
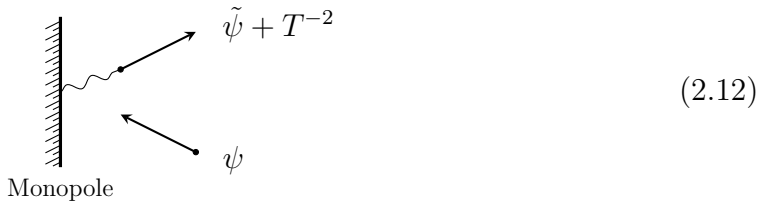
\begin{equation}
\begin{tikzpicture}[baseline=.9cm]
 
\draw[very thick] (-3,0) -- (-3,2);
\draw[->,>=stealth,thick] (-1.2, 0.2) -- (-2.2, 0.7);
\draw[<-,>=stealth,thick] (-1.2, 1.8) -- (-2.2, 1.3);
\draw[decorate, decoration={snake,amplitude=.6mm}] (-3,1) -- (-2.2, 1.3);
 
\foreach \x in {0,...,12} \draw (-3-.2,.15*\x+.04) -- (.2-3-.2,.15*\x+.1+.04);
\foreach \x in {-1,...,11} \draw (-3-.1,.15*\x+.09+.075) -- (.2-3-.2,.15*\x+.1+.04+.075);
 
\node at (-.1, 1.9) {$\tilde{\psi} + T^{-2}$};
\node at (-1.7+1,.1) {$\psi$};
 
\filldraw (-1.2, 0.2) circle (1pt);
\filldraw (-2.2, 1.3) circle (1pt);

\node[scale=.8] at (-3, -.3) {Monopole};

\end{tikzpicture}
\end{equation}
This topological line acts as a branch cut. In Euclidean space, it has the property that when the local operator $\tilde{\psi}$ passes through the line, it picks up a phase. The  fermion $\psi$ is left untouched. Said differently, the fermion $\tilde{\psi}$ picks up a monodromy $\omega$ when dragged around the twist operator $T(x)$. For the scattering, we need the topological line $T^{-2}$, under  which  $\tilde{\psi}$ picks up the monodromy $\omega^{-2}$. In~\eqref{twist} we abuse notation by using $T^{-2}$ both to denote the topological line and its endpoint.


\para
As is clear from the picture, the topological line cuts spacetime in two. There is before, and there is after. That makes it difficult to observe the line operator in Minkowski space by performing a suitable Aharonov-Bohm measurement. However there is a more dramatic effect because the twist operator carries charges. In the present case, $T^{-2}$ can be shown to carry $U(1)$ gauge charge $+2$. This is what ensures that charge is conserved in the scattering. We will review the calculation of the charge of $T$ in section~\ref{twistsec} below. 

\para
Viewed from the perspective of 4d scattering, the topological line blossoms into a 3d topological surface. It can be viewed as filling the outgoing lightcone, as shown below. However, precisely because the surface is topological, the apex of the cone can move up and down the monopole end point with impunity. 
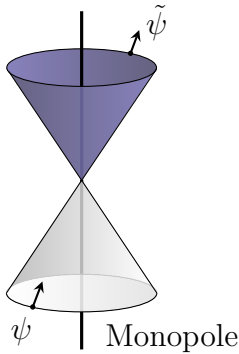
\be
\begin{tikzpicture}[scale=1/2]
\draw[very thick] (0,-7.5) -- (0,-2.99);
\fill[left color={rgb,255:red,75;green,69;blue,129},right color={rgb,255:red,144;green,133;blue,250},opacity=.5] (2,0) -- (0,-3) -- (-2,0) arc (180:360:2cm and 0.5cm);
\fill[left color=gray!50!white,right color=white,opacity=1] (2,-6) -- (0,-3) -- (-2,-6) arc (180:360:2cm and 0.5cm);
\draw[very thick,color={rgb,255:red,75;green,69;blue,129},opacity=.4] (0,-3) -- (0,-.5);
\draw[very thick,color=black,opacity=.1] (0,-3) -- (0,-6.5);
\fill[right color={rgb,255:red,75;green,69;blue,129},left color={rgb,255:red,144;green,133;blue,250},opacity=.5] (0,0) circle (1.99cm and 0.5cm);
\draw[very thick] (0,-.5) -- (0,1.5);
\draw (-1.94,-.1) -- (0,-3) -- (1.99,-.04);
\fill[right color=gray!20!white,left color=white] (0,-6) circle (1.99cm and 0.5cm);
\draw[very thick,color=black,opacity=.1] (0,-3) -- (0,-6.5);
\draw (0,0) circle (1.99cm and 0.5cm);
\draw (-2,-6) arc (180:360:2cm and 0.5cm);
\draw[black!20!white] (-2,-6) arc (180:0:2cm and 0.5cm);
\draw (-2,-6) -- (0,-3) -- (2,-6);
\fill (-1.3,-6.37) circle (2pt); \draw[thick,->,>=stealth] (-1.3,-6.37) -- (-1.3+.258,-6.37+.68);
\fill (1.3,.38) circle (2pt); \draw[thick,->,>=stealth] (1.3,.38) -- (1.53,1);
\node at (2.4,-7.2) {Monopole};
\node at (-1.6,-7) {${\psi}$};
\node at (2,1.2) {$\tilde{\psi}$};
\end{tikzpicture}
\nn\ee

\noindent
Scattering off monopoles with $g>1$ in this theory brings a new ingredient, namely the need for a non-invertible symmetry and, following \cite{Choi:2022jqy,Cordova:2022ieu}, the associated ABJ defect \cite{vanBeest:2023dbu}. We will not have need for this more elaborate construction in this paper. 

\para
Before we review the properties of these twist operators, we first give one further example of scattering. This is a case where we believe that consistent scattering should be possible but we do not know how to accomplish it. The example is a simple $U(1)$ chiral gauge theory, with five left-handed fermions with charges $q_i=-1,-5,7,8,-9$. These charges are designed so that the gauge anomaly vanishes, $\sum q_i^3=0$, and the mixed gauge-gravitational anomaly vanishes, $\sum q_i=0$. For a monopole with $g=1$, the chiral ingoing and outgoing fermions transform under $U(1)\times SU(2)_\text{rot}$ as
\begin{equation}
\begin{array}{l|cc}
&U(1) &SU(2)_\text{rot}\\\hline
\text{Incoming:} & +7 & {\bf 7} \\ & +8 & {\bf 8}   \\ \hline
\text{Outgoing:}\quad& +1 & {\bf 1} \\ & +5 & {\bf 5} \\ & +9 & 
{\bf 9}
\end{array}
\label{dunno}\end{equation}
%
%
This time the 2d anomalies cancel. We have $c_L=c_R$ and $\sum\gamma^\star_i q_i^2=0$ where $\gamma^\star_i=\pm 1$ for ingoing/outgoing fermions respectively. Indeed, it is straightforward to show that if the 4d gauge and gravitational anomalies cancel, then so too do the 2d anomalies \cite{Cordova:2018cvg,Maldacena:2020skw, vanBeest:2023dbu}. 

\para
Given that the anomalies cancel, there is a general expectation that it should be possible to implement a boundary condition for the fermions in \eqn{dunno} preserving both $SU(2)_\text{rot}$ and $U(1)$. But no such boundary condition is known. Relatedly, it is not known what defect operator, if any, should be attached to the outgoing radiation to make scattering consistent. These remain open questions.

\para
This  leads us to ask: when can we find consistent scattering data, and when do we fail? This question is most naturally answered from the perspective of 2d boundary conformal field theory. Suitable boundary conditions are typically known only when the CFT is rational with respect to the symmetry we wish to preserve. For us, that means that the number of ingoing (or outgoing) 2d Weyl fermions is equal to the rank of the symmetry group. Indeed, it was previously noted in \cite{Callan:1983ed} that in this situation, one can bosonise the fermions and impose simple, linear boundary conditions on the periodic scalars.  More generally, boundary conditions for   $N$ free fermions preserving chiral, non-anomalous $U(1)^N$ symmetries were constructed in \cite{Smith:2019jnh,Smith:2020rru}.  But when the number of fermions exceeds the rank of the symmetry group, we are usually all at sea. It's not necessarily that suitable boundary states don't exist. It's just that we typically don't know how to construct them.

\para
We can see this at play in the examples above. In the theory \eqn{sun} with $SU(N)\times U(1)$ symmetry, we had $N$ ingoing and $N$ outgoing fermions. That's why we could make progress.  However, in the theory \eqn{dunno}, with just $SU(2)\times U(1)$ symmetry, we have 15 ingoing and outgoing fermions. 

\para
There are exceptions to this statement. For the gapless QED \eqn{sqed} with a monopole of higher charge $g>1$, the number of ingoing fermions exceeds the number of symmetries. Nonetheless, by imposing a few reasonable further constraints, namely that the symmetry defect commutes with $SU(N) \times SU(2)_\text{rot}$, and is purely right-moving in 2d, a candidate for the defect operator was proposed in  \cite{vanBeest:2023dbu}. As general rule of thumb, if the number of fermions is equal to the rank of the symmetry group or one less, then the boundary state is known. Otherwise, one has to be lucky or clever or both. 
The distinction between theories where we can solve the scattering and those where we can't will be important when we come to discuss the Standard Model.

\subsection{Twist Operators}\label{twistsec}

The twist operator is the key to monopole scattering. These objects are well understood in two-dimensional conformal field theories where they appear as a local operator, trailed by a monodromy line. The properties of these operators, such as their dimension and their charge,  can be determined by invoking the  state-operator map. 

\para
Although our scattering set-up takes place on a manifold with boundary, we are ultimately interested in local properties of operators, such as their dimension and charge and, for this, it will suffice to consider the 2d CFT on $\R^2$. Local operators on $\R^2$ are mapped to states  on $\R\times {\bf S}^1$. This map can be extended to twist operators which don't obviously qualify as ``local'' because of the trailing branch cut (such operators are sometimes called \emph{point} operators, since they are defined at a point but are not strictly local; see~\cite{Shao:2023gho} for a recent review).

\para
To keep things reasonably general, consider a $\Z_N$ twist operator $T$ placed at the origin of $\R^2$. This operator is defined by its action on local operators in the theory as they move around the origin. For our purposes, we will restrict to theories of free fermions $\psi_i$. When each of these fermion operators is dragged around the origin, it picks up a phase
\be \psi_i \ra \omega^{q_i}\psi_i\ \ \ \text{with}\ \ \omega = e^{2\pi i /N}.\ee
After a conformal transformation to the cylinder, this monodromy affects the boundary conditions of the fermions. We denote the coordinate on ${\bf S}^1$ as $\sigma \in [0,2\pi)$. The presence of the twist operator means that we should take boundary conditions
\be \psi_i(\sigma + 2\pi)  = - \omega^{q_i} \psi_i(\sigma)\ .\label{twistbc}\ee
Here the overall minus sign reflects that fact that, in the absence of any twist operator, we have NS boundary conditions on the circle. If we work in Lorentzian signature, then quantising the theory with boundary conditions \eqn{twistbc} gives rise to a different, twisted Hilbert space. For all  applications of monopole scattering in this paper,  the $\Z_N$ twist acts only on the outgoing fermions. This is the origin of the statement that the  monopole is a portal to a twisted Hilbert space.

\para
The phrase ``twisted Hilbert space'' is, strictly, not accurate for the problem of monopole scattering. The idea of a twisted Hilbert space really holds only when space is compact and a timelike defect is inserted, as above. Clearly, however, for monopole scattering space is non-compact. Nonetheless, the outgoing state is not in the conventional Fock space of the theory, at least if defined by acting with finitely many local operators. Instead, the outgoing state is a very complicated excitation in the untwisted Hilbert space, but one which has an economical description in terms of the spacelike (or null) topological surface defect. 

\para
Like all local operators in a CFT, the twist operator has a charge and dimension. Both are simple to compute using the state-operator correspondence. For a given field $\psi$, the NS-NS partition function, augmented by twists of $\psi \rightarrow e^{2 \pi i \eta_{t,x}} \psi$ around the temporal and spatial cycles, is given by
\be \mathcal{Z}[\eta_x,\eta_t] = \frac{1}{\eta(\tau)} \sum_{n \in \mathbb{Z} - \eta_x} q^{n^2 / 2} e^{2 \pi i \eta_t n} \quad \text{with } q = e^{2 \pi i \tau}.\label{partfn}\ee
The expression \eqref{partfn} counts states with twisted boundary conditions $\psi(\sigma + 2 \pi) = -e^{2 \pi i \eta_x} \psi(\sigma)$, with each state weighted by $q^h e^{2 \pi i \eta_t Q}$ where $h$ is the dimension and $Q$ is the fermion number. From this it is straightforward to read off the properties of the vacuum state. The ground state minimises $n^2$ with $n\in \Z-\eta_x$, and so corresponds to
\be n = [\tfrac{1}{2} - \eta_x] - \tfrac{1}{2} \label{nvac}\ .\ee
Here $[x] \coloneqq x - \lfloor x \rfloor$ is the fractional part of $x$. The ground state then has dimension and fermion number
\be h = \frac{1}{2}n^2 \  \  \ \text{and}\ \ \  Q = n \label{dimcharge}\ee
As expected, both of these  vanish if we take $\eta_x = 0$ so that we have NS boundary conditions. The case with $\eta_x = 1/2$ corresponds to Ramond boundary conditions, in which case there are two degenerate vacua with $n = \pm\frac{1}{2}$. Here we will continue to call the choice of state $n = -\frac{1}{2}$ given by \eqref{nvac} ``the vacuum'', although it is degenerate with the state labelled by $n + 1=+\frac{1}{2}$. We'll say more about these zero modes below because they are relevant for the Standard Model.

\para
From \eqn{dimcharge}, we can compute the $U(1)$ charge of the ground state in the twisted sector. Each fermion has a twist $(\eta_x)_i = q_i/N$.  We then sum over all fermions, weighted with their $U(1)$ charge $Q_i$,
\be Q_\text{vac} = \sum_i Q_i \left(\left[\frac{1}{2} - \frac{q_i}{N}\right] - \frac{1}{2}\right) \gamma^\star_i \label{charge}\ee
Here $\gamma_i^\star=\pm 1$ for left-moving and right-moving fermions respectively. This is the charge of the twist operator $T$ that sits at the end of the topological line,
\be Q[T] = Q_\text{vac}\ .\ee
The computation of the dimension of the twist operator proceeds in a similar way. The dimension of the twist operator is $h[T] = h_\text{vac}$ with
\be h_\text{vac} = \frac{1}{2} \sum_i \left(\left[\frac{1}{2} - \frac{q_i}{N}\right] - \frac{1}{2}\right)^2 \label{hvac}\ee
The scattering process results in a twist operator together with some number of fermion excitations whose contribution to the overall dimension should also be included. For the boundary condition $\psi(\sigma + 2\pi) = -e^{2\pi i \eta_x} \psi(\sigma)$, the fermion has dimension $h[\psi] = [\frac{1}{2} - \eta_x]$. This agrees with the usual dimension $h[\psi] = 1/2$ for a fermion with NS boundary conditions, and captures the fact that the fermion has a zero mode, with $h[\psi]=0$, for Ramond boundary conditions. To illustrate this, we turn to an example.

\subsubsection{An Example: Scattering Preserving $U(1)\times SU(N)$}

Following \cite{vanBeest:2023dbu}, we now apply these ideas to monopole scattering in the theory  \eqn{sqed}, consisting of a  $U(1)$ gauge theory with an $SU(N)$ flavour symmetry. The fermions $\psi$ transform in the ${\bf N}_{+1}$ and fermions $\tilde{\psi}$ in the ${\bf N}_{-1}$.  This means that, in the presence of a monopole of charge $g=1$, the s-wave consists of ingoing modes transforming in the  ${\bf N}_{+1}$ and outgoing modes in the ${\bf N}_{-1}$. 

\para
As we anticipated in \eqn{twist}, the outcome of the scattering involves a twist operator
\be \psi \longrightarrow \tilde{\psi} + T^{-2}\label{lastsummer}\ee
where $T$ is the generator of a $\Z_N$ symmetry that does not act on $\psi$, but acts on the outgoing fermions as $T:\tilde{\psi} \ra e^{2\pi i /N} \tilde{\psi}$, and so $T^{-2}: \tilde{\psi} \ra e^{-4\pi i /N} \tilde{\psi}$.\footnote{For even $N$, the symmetry generated by $T$ actually suffers a non-perturbative anomaly, while only $T^2$ is anomaly-free. However, for simplicity, we will continue to write all generators in terms of $T$.} We can use the formulae above to compute the charge and dimension of the twist operator. To this end, we work with the Cartan of the $U(N)$ symmetry, under which the fermions $\psi_i$ have charges $Q_{\alpha i}$ and the fermions $\tilde{\psi}_i$ have charges $\tilde{Q}_{\alpha i}$, where $\alpha =1,\ldots,N$ labels the $U(1)$ group and $i=1,\ldots, N$ labels the fermion. These charge matrices take the form
%
%
\be  Q = \left(\begin{smallmatrix}+ & + & + & \ldots & + \\ + & - &  &&\\ & + & - && \\  & & & \ddots & \\ & & & & -\end{smallmatrix}\right)\ \ \ \text{and}\ \ \ \tilde{Q} = \left(\begin{smallmatrix} - & - & - & \ldots & - \\ + & - &  &&\\ & + & - && \\  & & & \ddots & \\ & & & & -\end{smallmatrix}\right)\ .\ee
Here the top row corresponds to the original $U(1)$, and the remaining rows are the charges under the Cartan $U(1)^{N-1} \subset SU(N)$. We can compute the charge of the twist operator $T^{-2}$ using our formula \eqn{charge}, with $q_i=-2$ for each fermion $\tilde{\psi}$. We will restrict to $N\geq 5$. (The case of $N=4$ involves fermion zero modes and will be dealt with separately below.) We have $\gamma_i^\star=-1$, because the $\Z_N$ symmetry acts only on outgoing fermions. We then find that the charges under $U(1)^N$ are given by
\be  Q_\alpha [T^{-2}] = - \sum_i \frac{2}{N}\,\tilde{Q}_{\alpha i}  =  (+2,0,0,\ldots,0) \ .\ee
That string of zeros is just telling us that the twist operator $T^{-2}$ is a singlet under $SU(N)$. But it has charge $+2$ under $U(1)$. That's what we wanted: it means that the scattering \eqn{lastsummer} preserves $U(1)\times SU(N)$ as advertised.

\para
We can similarly compute the dimension of the twist operator.  From \eqn{hvac}, we get $h[T^{-2}] = 2/N$. Meanwhile, we have 
\be h[\tilde{\psi}] = \frac{1}{2}-\frac{2}{N}\ \ \ \Longrightarrow\ \ \ h[\tilde{\psi} + T^{-2}] = \frac{1}{2}\ .\ee
We see that the combination of the outgoing fermion, together with the twist operator, has the dimension appropriate to a free fermion in two dimensions. 

\subsubsection{Zero Modes}\label{zerosec}

The case of $N=4$ deserves special attention because there are fermion zero modes. This case will also arise when we discuss the minimal monopole in the Standard Model so it is worth elaborating on in detail.

\para
First consider a single Weyl fermion $\psi$ on ${\bf S}^1$ with Ramond boundary conditions and charge $Q$ under a $U(1)$ global symmetry. In this case, the ground state is not unique. Instead there is a complex fermion zero mode, resulting in a degeneracy of ground states $|{0}\ket$ and $\psi|{0}\ket$. The two states have a charge that differs by $Q$, and CPT invariance requires that they are given by
\be Q_\text{vac} = \pm \frac{Q}{2}\ .\ee
This fact is captured by the partition function \eqref{partfn}, with the highest-weight states labelled by $n = -\ft12$ and $n = \ft12$ corresponding to $|{0}\ket$ and $\psi|{0}\ket$ respectively, and their $U(1)$ charges given by \eqref{dimcharge} as $nQ$.

\para
For our monopole scattering problem, we have four  fermion zero modes  $\tilde{\psi}_i$, $i=1,2,3,4$, giving rise to $2^4$ degenerate ground states when the theory is quantised on a circle. These sit in the representations:
\begin{equation}
\begin{array}{c|cc}
&U(1) &SU(N)\\\hline
|0\ket \ \ \  & +2 & {\bf 1}\\
\tilde{\psi}|0\ket \ \ \  & +1  & {\bf 4}\\
\tilde{\psi}^2 |0\ket \ \ \ & 0 & {\bf 6} \\
\tilde{\psi}^3|0\ket \ \ \ &  -1 & \bar{\bf 4} \\
\tilde{\psi}^4|0\ket\ \ \  & -2 & {\bf 1} 
\end{array}
\label{table}\end{equation}
%
The operators associated to each of these states have dimension $1/2$, the same as a free fermion. Now we see what happens under scattering: if we throw in $\psi$ in the ${\bf 4}_{+1}$, we simply get out the twist operator in the state ${\bf 4}_{+1}$. This can be viewed as the general result \eqn{lastsummer}, just with the fermion $\tilde{\psi}$ acting as a zero mode.

\para
Viewed from a purely two-dimensional perspective, this coincides with the problem studied by Maldacena and Ludwig long ago \cite{Maldacena:1995pq}. Ignoring the complex structure imposed by $SU(4)$, one could instead view this as a system with $\mathrm{Spin}(8)$ symmetry, in which case the zero modes furnish the well-known spinor representations ${\bf 8}_s\oplus {\bf 8}_c$. Our proposal here is then in agreement with Maldacena and Ludwig's. This boundary condition has been implemented experimentally in the guise of the two channel Kondo problem \cite{Potok_2007}.

\section{Monopoles in the Standard Model}\label{monosec}

We now turn to monopoles in the Standard Model. Neglecting global issues, the gauge group is, of course, $SU(3) \times SU(2) \times U(1)$. However, when it comes to monopoles, global issues are important. For us, it will be important that we take the gauge group to be
\be  G = \frac{SU(3)_S\times SU(2)_W \times U(1)_Y}{\Z_6}\ .\label{G}\ee
The group $G$ acts faithfully on the matter multiplets, which sit in the following representations:
\begin{equation}
\begin{array}{lc|ccc}
&&SU(3)&SU(2)&U(1)_Y\\\hline
\text{Left-Handed:}\quad&l_L&{\bf 1}&  {\bf 2}& -3\\
&q_L&  {\bf 3} & {\bf 2} & +1 \\ \hline
\text{Right-Handed:} \quad & e_R & {\bf 1} & {\bf 1} & -6 \\
& u_R & {\bf 3} & {\bf 1} & +4 \\
& d_R & {\bf 3} & {\bf 1} & -2 
\end{array}
\label{sm}\end{equation}
%
%
We've normalised the $U(1)_Y$ hypercharge to be integer-valued, rather than the more traditional fractional value (e.g.\ the left-handed quarks are usually taken to have hypercharge $1/6$). We will also have cause to work with conjugated fields, as indeed we did in \eqn{puzzle}. So, for example, $\bar{u}_R$ is a left-handed Weyl fermion in the representation $(\bar{\bf 3},{\bf 1})_{-4}$.

\para
Each of these representations is untouched by the $\Z_6$ quotient which is generated by
\be  ( e^{2\pi i /3}\mathds{1}_3 , -\mathds{1}_2\ , e^{2\pi i /6})\in SU(3)_S \times SU(2)_W \times U(1)_Y\  .\label{z6}\ee
The Higgs field transforms in the $({\bf 1},{\bf 2})_3$ representation and is similarly uncharged under $\Z_6$.

\para
In addition to these gauge symmetries, the classical Lagrangian admits a $U(1)^4$ global symmetry, amongst which we find the baryon and lepton number symmetries that are preserved if we turn on the usual Yukawa couplings.\footnote{See~\cite{Cordova:2022qtz} for a careful analysis of the symmetries of the Standard Model, including higher categorical and non-invertible symmetries.} Only the subgroup $U(1)_{B-L}\subset U(1)^4$, under which quarks $q_L,u_R,d_R$ have charge $+\frac{1}{3}$ and leptons $l_L,e_R$ have charge $-1$, is free of an ABJ anomaly. Outside this subgroup, we have mixed anomalies\footnote{There is a unique $U(1)_e\subset U(1)^4$ independent from $U(1)_{B-L}$ which has a mixed anomaly only with $U(1)_Y$. This $U(1)_e$ acts only on $e_R$. Discrete rotations in $U(1)_e$ can be recovered as non-invertible symmetries. We won't need these for our scattering, however.} with $SU(3)_S$, $SU(2)_W$ and $U(1)_Y$.  Only a $(G\times U(1)_{B-L})/\Z_3$ acts faithfully on the matter content, with the quotient acting as\footnote{Note, here we use proper normalisation for $U(1)_{B-L}$, e.g.\ under $e^{i\theta}\in U(1)_{B-L}$, $l_L\to e^{-3i\theta}l_L$. } $(e^{2\pi i/3},e^{-2\pi i/3})\in U(1)_{B-L}\times U(1)_Y$. Note also that $(-1)^F=-1\in U(1)_{B-L}$.



 
\para
 The fact that we quotient the gauge group by $\Z_6$ means that we restrict the allowed electric representations of the theory. So the gauge group $G$ does not, for example, allow for the introduction of quark matter that is neutral under hypercharge. However, as explained in \cite{Aharony:2013hda}, restricting the electric representations allows for more magnetic representations consistent with Dirac quantisation. That means that the gauge group $G$, with the $\Z_6$ quotient, affords more options for 't Hooft lines  \cite{Tong:2017oea}. 
 
\para
The 't Hooft lines are implemented by cutting out a ball $B^3$ around the spatial origin and imposing boundary conditions on the resulting ${\bf S}^2 = \partial B^3$. For our purposes, it's important that these boundary conditions preserve rotational invariance so that it makes sense to decompose the fermions in monopole harmonics. 

\para
The most general such monopole is parameterised by four integers, one for each Cartan element of $G$. These are analogous to the GNO charges of \cite{Goddard:1976qe}.  (Analogous, but not quite the same as in the original GNO paper, where the gauge group was spontaneously broken to its Cartan by an adjoint scalar and the monopoles were solitons. A clean discussion of the situation for 't Hooft lines can be found, for example, in \cite{Dyer:2013fja}.) We denote these integers as $n_1$ and $n_2$ for the  $SU(3)$ magnetic fluxes, as $m$ for the $SU(2)$ flux, and as $g$ for the $U(1)_Y$ flux. After invoking a gauge transformation to rotate the fluxes into the diagonal, Cartan elements, the most general rotationally invariant magnetic monopole takes the form (in a slight abuse of notation)
\be
  F = \begin{pmatrix}
  	n_1 - \frac{N}{3}	& &		\\
  	 & n_2 - \frac{N}{3}		&	\\
  	 &&  - \frac{N}{3}
  \end{pmatrix}\oplus\begin{pmatrix}
  	\frac{1}{2}m		&		\\
  	&	-\frac{1}{2}m
  \end{pmatrix}\oplus  
  \left(g+\frac{N}{3} + \frac{m}{2}\right) \left(\frac{1}{2} \sin \theta \,d \theta \wedge d \phi\right)
\nn\ee
Here $N=n_1+n_2$, and we can use the Weyl group of $SU(3)\times SU(2)$ to insist that $n_1\geq n_2\geq 0$ and $m\geq 0$. 

\para
The existence of the $\Z_6$ quotient in \eqn{G} allows us to take $N,m\in \Z$. In the absence of this quotient, we must have $N\in 3\Z$ and $m\in 2\Z$.

\para
The characterisation of the monopole in terms of four integers is finer than a crude topological characterisation~\cite{Lubkin:1963zz}. The latter is given by $\pi_1(G) = \Z$ and is labelled by the combination $p=2N+3m+6g$. Said another way, the theory has a $U(1)$ magnetic 1-form symmetry; the charge of the `t Hooft line under this 1-form symmetry is the integer $p$. This means that monopoles that lie in the same topological class can be deformed into each other by moving away from the rotationally invariant ansatz. 
Of course, the 't Hooft lines described here are really a proxy for some dynamical monopoles that are expected to arise in a UV completion of the Standard Model. From this low-energy perspective, we treat the $(n_1,n_2,m,g)$ as the conserved charges.

\para
For much of this paper, we will neglect gauge field fluctuations. However, when such effects are included, magnetic charge can be screened, and a generic $(n_1,n_2,m,g)$ monopole is unstable to W-boson condensation \cite{Aharony:2022ntz,Aharony:2023amq}. The only data that is robust under such effects is precisely the 1-form charge $p$. Indeed, we find by explicit computation that all monopoles exhibit this instability, except those with $n_1,n_2,m\le 1$. It follows that for each value of the 1-form magnetic charge $p$, there is precisely  \textit{one} stable monopole. Details of this computation can be found in Appendix~\ref{stability}.

\para
For $n_i,m\neq 0$, the 't Hooft line breaks the gauge group $G$. The form of this breaking depends on the fluxes. 
\begin{itemize}
\item For generic fluxes, the group $G$ is broken to the maximal torus
\be G\longrightarrow H=\frac{U(1)_S^2 \times U(1)_W \times U(1)_Y}{\Z_6}\ .\label{fullbreaking}\ee
\item
We will  be particularly interested in situations where  either $n_1=n_2\neq 0$ or where $n_1\neq 0$ and $n_2=0$. For both of these choices of flux, with $m\neq 0$, we have
\be
 G\longrightarrow H=\frac{SU(2)_S\times U(1)_S \times U(1)_W \times U(1)_Y}{\Z_6}\ .\label{littlebreaking}\ee
\item If $n_1=n_2=0$ then $SU(3)_S\subset G$ is left unbroken. If $m=0$ then $SU(2)_W\in G$ is left unbroken. 
\end{itemize}
In the rest of this section, we pose the question of fermion scattering off different classes of monopoles. This means that we decompose each of the Standard Model fermions \eqn{sm} into monopole harmonics and determine the ingoing and outgoing modes in the lowest angular momentum sector. We will do this for the generic monopole, with breaking \eqn{fullbreaking}, and for a particular  monopole with breaking \eqn{littlebreaking}. (We do not present results for the special cases with unbroken $SU(3)_S$ or $SU(2)_W$, although it is straightforward to do so. There are no surprises.)

\subsection{The Minimal Monopole}

We start by describing the situation for the minimal monopole. This is ``minimal'' in the sense that the topological charge is $2N+3m+6g=1$ and is achieved by taking the fluxes
\be n_1 = n_2 =1\ \ ,\ \ m=1\ \ ,\ \ g=-1\ .\label{minflux}\ee
Although this choice doesn't look particularly natural, this is the combination that gives what you would normally call \emph{the} magnetic monopole, meaning that after electroweak symmetry breaking it has minimal Dirac charge with respect to the electron. The fact that $n_i\neq 0$ means that this monopole necessarily carries colour magnetic charge, and this is what ensures Dirac compatibility with fractionally charged quarks. 

\para
The fluxes \eqn{minflux} are such that the symmetry breaking structure is \eqn{littlebreaking}, with a surviving $SU(2)_S\times U(1)_S\subset SU(3)_S$. The $U(1)_S$ is generated by 
\be \begin{pmatrix}  1 & & \\ & 1 & \\ & & -2\end{pmatrix}\ .\ee
%
%
For each fermion in \eqn{sm}, we must figure out the quantum numbers under the unbroken $H$, together with the lowest angular momentum mode. The latter is determined by an appropriate combination of the magnetic  flux multiplied by the charge. Quarks with colour in the $a=1,2$ components experience flux $-1/3$ while those with $a=3$ experience $+2/3$. Weak doublets experience flux $\pm 1/2$. Finally all fermions experience
\be \left(g+\frac{N}{3} + \frac{m}{2}\right)Y = \frac{Y}{6} \ee
from their hypercharge. These contributions are then added and the result, $w\in \Z$, contains the relevant information about the lowest angular momentum mode. Left-handed fermions transform in the $|w|$-dimensional representation of $SU(2)_\text{rot}$ and are ingoing if $w>0$ and outgoing if $w<0$.  This is inverted for right-handed fermions. 

\para
Something rather special happens for this minimal monopole: there are a number of components of fermions for which $w=0$. This means that they are blind to the monopole and have trivial scattering. This happens for the left-handed fermions $l_L^+$,  $q_L^{a-}$, and $q_L^{3+}$ where $a=1,2$ is the colour index and the $\pm$ superscript denotes the two components of the $SU(2)_W$ doublet. It also happens for the right-handed  fermions $u_R^3$ and $d_R^a$ for $a=1,2$. 

\para
In addition, all other fermions have $w=\pm 1$ which means that they are in the s-wave. The upshot is that we have 4 ingoing and 4 outgoing fermions, transforming under $SU(2)_S\times U(1)_S\times U(1)_W\times U(1)_Y$ as
\begin{equation}
\begin{array}{lc|cccc}
&&SU(2)_S&U(1)_S&U(1)_W& U(1)_Y\\\hline
\text{Incoming:}\quad&q^{a+}_L&\boldsymbol 2&1&1& 1\\
&d^3_R&\boldsymbol1 & -2 & 0 & -2\\
&e_R&{\boldsymbol 1}&0 & 0 & -6\\\hline
\text{Outgoing:}\quad& u_R^a&\boldsymbol2& 1 & 0 & 4\\
&q_L^{3-}&\boldsymbol1&-2 & -1 & 1\\
&l_L^-&{\boldsymbol1}&0 & -1 & -3
\end{array}\label{cando}
\end{equation}
%
%
This was the problem originally considered by Callan in \cite{doi:10.1063/1.34591}. We see that we have just 4 ingoing fermions, and require boundary conditions that preserve a group $H$ of rank 4. This, therefore, falls within the class of scattering problems that can be solved. Indeed, as we will see, it's closely related to the $SU(4)$ theory (and to the two-channel Kondo problem!) that we described in the previous section. We will turn to the scattering in section~\ref{nicesec}. 

\subsubsection*{The Anti-Monopole}

For completeness, we also look at the  minimal anti-monopole that has total topological charge $2N+3m+g=-1$. This arises from the fluxes
\be n_1=1\ \ ,\ \ n_2=0\ \ ,\ \ m=1\ \ ,\ \ g=-1\ .\ee
The lack of symmetry with \eqn{minflux} is entirely due to our choice of parameterisation and, in particular, the use of the Weyl group to set $n_1\geq n_2\geq 0$. This time, the ingoing and outgoing modes are flipped, with
%

\begin{equation}
\begin{array}{lc|cccc}
&&SU(2)_S&U(1)_S&U(1)_W& U(1)_Y\\\hline
\text{Incoming:}\quad&q^{1+}_L&\boldsymbol 1&2&1& 1\\
&u^a_R&\boldsymbol2 & -1 & 0 & 4\\
&l_L^+&{\boldsymbol 1}&0 & 1 & -3\\\hline
\text{Outgoing:}\quad& d_R^1&\boldsymbol1& 2 & 0 & -2\\
&q_L^{a-}&\boldsymbol2&-1 & -1 & 1\\
&e_R&{\boldsymbol1}&0 & 0 & -6
\end{array}\label{anticando}
\end{equation}
%
Now the doublet of colour indices range over $a=2,3$. This, and other, minor differences from \eqn{cando} can be traced to Weyl reorderings, arising from our choice of parameterisation of the fluxes. 

\subsection{The Generic Monopole}\label{genericsec}

As we've seen, the minimal monopole is transparent to 7 of the 15 fermions in one generation of the Standard Model. This will allow us to solve for the scattering in section~\ref{nicesec}. For the generic monopole, there is no such simplification.

\para
To illustrate this, here we give the fermion decomposition for a general monopole with $n_1\geq n_2\geq 0$, $m\geq 0$ and any $g\in \Z$. The monopole generically breaks the gauge group to the Cartan \eqn{fullbreaking} and, as we have seen, there can be symmetry enhancement for special cases.  Each fermion then decomposes into a bunch of representations of 
\be SU(2)_\text{rot} \times U(1)_S^2 \times U(1)_W \times U(1)_Y\ .\ee
We will take the generators of $U(1)_S^2$ to be $\diag(1,-1,0)$ and $\diag(0,1,-1)$ and the generator of $U(1)_W$ to be $\diag(1,-1)$. We then decompose each fermion into representations of $U(1)_S^2\times U(1)_W\times U(1)_Y$, and determine the total flux times charge $w$ experienced by each such component. We use the notation $[w]$ for each component, where two pieces of information are stored in the integer $w$: the lowest angular momentum mode is in the $|w|$-dimensional representation of $SU(2)_\text{rot}$, and this mode is ingoing (outgoing) for $w>0$ ($w<0$). If $w=0$, the fermion is blind to the monopole. We have the following decomposition
\be 
l_L &\longrightarrow& [-3g-n_1-n_2-m]_{0,0,1,-3}\oplus [-3g-n_1-n_2-2m]_{0,0,-1,-3} \nn\\
q_L &\longrightarrow& [g+n_1+m]_{1,0,1,1}\oplus [g+n_1]_{1,0,-1,1} \oplus [g+n_2]_{-1,1,-1,1}\nn \\ 
&& \oplus\ [g+n_2+m]_{-1,1,1,1} \oplus [g+m]_{0,-1,1,1}\oplus [g]_{0,-1,-1,1}\nn\\
e_R &\longrightarrow& [-6g-2n_1-2n_2-3m]_{0,0,0,-6} \nn\\
u_R &\longrightarrow& [4g+2n_1+n_2+2m]_{1,0,0,4} \oplus [4g+n_1+2n_2+2m]_{-1,1,0,4} \nn\\ &&\oplus\ [4g+n_1+n_2+2m]_{0,-1,0,4}
\nn\\d_R &\longrightarrow& [-2g-n_2-m]_{1,0,0,-2} \oplus [-2g-n_1-m]_{-1,1,0,-2} \nn\\ &&\oplus\ [-2g-n_1-n_2-m]_{0,-1,0,-2}
\ .\ee
By construction, we have $n_1\geq n_2\geq 0$ and $m\geq 0$, while $g$ can take any sign. If, in addition, we take $g\geq 0$ then the ingoing modes are all components of $q_L$, $e_R$ and $d_R$, while the outgoing modes are all components of $l_L$ and $u_R$. In this case, the number of ingoing modes equals the number of outgoing modes and is given by
\be c_L=c_R = 3(6g+2n_1+2n_2 +3m)\ .\ee
We recognise the quantity in brackets as the topological charge of the monopole. One can similarly check that the (2d)  't Hooft anomalies for $SU(2)_\text{rot} \times U(1)^4$ all cancel, as indeed they should if there is to be consistent scattering from ingoing  to outgoing states. 

\para
For $g<0$, we still have $c_L=c_R$, but for some values of $n_1,n_2,m$, a given fermion can have a combination of ingoing and outgoing modes.   Indeed, this happens for the minimal monopole where, as we have seen, the left-handed quarks $q_L$ contribute both ingoing and outgoing modes. The 't Hooft anomalies again cancel in this case. (The 't Hooft anomalies also cancel in situations where there is symmetry enhancement.)

\para
The symmetry group preserved by the monopole is made up of whatever of the gauge group is unbroken, along with $SU(2)_\text{rot}$ and $U(1)_{B-L}$. The fermion-monopole scattering problem for a generic monopole has $c_L > 6$, the rank of the preserved symmetry group. In addition, the $SU(2)_\text{rot}$ symmetry acts on the ingoing and outgoing modes in a complicated, chiral fashion. This means that nearly all monopoles in the Standard Model  fall firmly within the class of problems that we cannot (yet!) solve. The exceptions are the minimal monopole which has $c_L=4$, along with a handful of other examples with $c_L\le 6$, which we describe in Appendix~\ref{solvesec}.

\section{Scattering Off the Minimal Monopole}\label{nicesec}

In this section, we describe the resolution to the scattering problem for the minimal monopole. We will take our ``low-energy'' scattering to be at energy scales $E \gg 1$ TeV. This means that we neglect confinement and electroweak symmetry breaking and are justified in taking all fermions to be massless. Conversely, we want to treat the monopole as a 't Hooft line with no internal degrees of freedom, which means that we should also restrict ourselves to energies $E\ll M_\text{mono}$ where $M_\text{mono}$ is the mass of the monopole.

\para
For the purposes of our immediate discussion, it will be simplest if we make the slightly awkward choice of taking $SU(3)_S\times SU(2)_W\times U(1)_Y$ to be a global symmetry, with the monopole a background gauge field. Clearly this is unphysical, but it makes the physics easier to highlight. We will then discuss how things are affected by dynamical gauge fields in section~\ref{whatsec}.

\para
Our interest is in the  minimal monopole, preserving  $SU(2)_S\times U(1)_S\times U(1)_W\times U(1)_Y$. As we saw in \eqn{cando}, the ingoing and outgoing modes carry quantum numbers 
%
%

\begin{equation}
\begin{array}{lc|cccc}
&&SU(2)_S&U(1)_S&U(1)_W& U(1)_Y\\\hline
\text{Incoming:}\quad&q^{a+}_L&\boldsymbol 2&1&1& 1\\
&d^3_R&\boldsymbol1 & -2 & 0 & -2\\
&e_R&{\boldsymbol 1}&0 & 0 & -6\\\hline
\text{Outgoing:}\quad& u_R^a&\boldsymbol2& 1 & 0 & 4\\
&q_L^{3-}&\boldsymbol1&-2 & -1 & 1\\
&l_L^-&{\boldsymbol1}&0 & -1 & -3
\end{array}\label{candodo}
\end{equation}
%
%
with $a=1,2$ the colour label. The question is, of course: how do ingoing states scatter into outgoing states? One, naive way to proceed is to work with the Cartan $U(1)_S^2 \times U(1)_W\times U(1)_Y$ and try to find linear combinations of outgoing states that match the quantum numbers of ingoing states. This is straightforward, and one finds the following scattering,
\be {q}_L^{1+}&\longrightarrow&\frac{1}{2}\left({u}^1_R + \bar{u}_R^2 + \bar{q}_L^{3-} + \bar{l}_L^-\right) \nn\\
{q}_L^{2+}&\longrightarrow&\frac{1}{2}\left(\bar{u}^1_R + {u}_R^2 + \bar{q}_L^{3-} + \bar{l}_L^-\right) \nn\\
d_R^3&\longrightarrow&\frac{1}{2}\left(\bar{u}^1_R + \bar{u}_R^2 + {q}_L^{3-} + \bar{l}_L^-\right) \\ \label{allwrong}
 e_R &\longrightarrow&\frac{1}{2}\left(\bar{u}^1_R + \bar{u}_R^2 + \bar{q}_L^{3-} + l_L^-\right)\ .\nn\ee
This is effectively  the puzzle uncovered in \cite{doi:10.1063/1.34591} that we advertised in the introduction. It is difficult to make sense of those factors of $1/2$.  (It is also difficult to find a way to preserve the full $SU(2)_S$, rather than just the Cartan, since the expressions above are not $SU(2)_S$ covariant.)

\para
The resolution, as described in section~\ref{knowsec}, is that the outgoing fermions emerge attached to a topological surface defect. In fact, the problem is essentially identical to the $U(1)\times SU(4)$ scattering that we considered previously, replete with zero modes for the twist operator.

\para
One can, however, find consistent $2\ra 2$ scattering off the monopole without having to bring twist states into the picture. For example, the following scattering process preserves all non-anomalous quantum numbers
\be q_L^{1+} +q_L^{2+} \ \longrightarrow\ \bar{q}_L^{3-}+ \bar{l}_L^-\ .\ee
This scattering violates baryon number and is at the heart of the Callan-Rubakov effect \cite{Rubakov:1982fp,Callan:1982au,Callan:1982ac,VARubakov_1988}. 

\para
To make contact with our previous results, we denote the ingoing and outgoing fermions as 
\be \psi_i = (q_L^{1+},q_L^{2+},d_R^3,e_R)\ \ \ \text{and}\ \ \ \tilde{\psi}_i = (q_L^{3-},u_R^1,u_R^2,l_L^-)\ .\label{allfermions}\ee
The charge matrices  $Q_{\alpha i}$ and $\tilde{Q}_{\alpha i}$ for these fermions under the Cartan of $U(1)_S^2\times U(1)_W\times U(1)_Y$ are
\be Q = \begin{pmatrix} 1 & -1 & 0 & 0 \\ 1 & 1 & -2 & 0 \\ 1 & 1 & 0 & 0 \\ 1 &1 & -2 & -6\end{pmatrix}\ \ \ \text{and}\ \ \ \tilde{Q}=\begin{pmatrix} 0 & 1 & -1 & 0 \\ -2 & 1 & 1 & 0 \\ -1 & 0 & 0 & -1 \\ 1 & 4 & 4 & -3\end{pmatrix}.\label{qandtq}\ee
The twist operator $T$ acts as a $\Z_2$ symmetry only on the outgoing fermions
\be T: \psi_i \mapsto \psi_i \ \ \ \text{and}\ \ \ T:\tilde{\psi}_i \mapsto -\tilde{\psi}_i\ .\label{2dtwist}\ee
We should identify this 2d topological line as the $S^2$ reduction of a topological surface in 4d.  This amounts to finding a $\Z_2 \subset G$. The following generator does the job:
 \begin{align}
  g &=  \left[\begin{pmatrix}
  	-1	&				&		\\
  				& -1	&		\\
  				&				& 1
  \end{pmatrix}\otimes\mathds{1}_2 \otimes (-1)\right]\ .\label{g1}
\end{align}
This acts on $\tilde{\psi}_i$ as $-1$ and on $\psi_i$ as $+1$. It also acts on some of the transparent fermions that don't play a role in the scattering.

\para
In fact, the choice of $\Z_2$ in \eqn{g1} is not unique. There is a $U(1)'\subset [U(1)_{B-L}\times G]/\Z_3$ under which all the fermions in \eqn{allfermions} are neutral, while those with $w=0$ that are blind to the monopole are not. This means that we can dress the $\Z_2$ action \eqn{g1} with an additional $\Z_2\in U(1)'$. The resulting action is 
\begin{align} g' &= (-1) \otimes \left[\mathds{1}_3\otimes\begin{pmatrix}
  	i 	& 		\\
  		& -i
  \end{pmatrix} \otimes (i)\right]
  \label{g2}
\end{align}
with the overall $(-1)\in U(1)_{B-L}$. While the first action \eqn{g1} lies entirely within the group $G$, this second action \eqn{g2} lies in $G\times U(1)_{B-L}$.

\para
As explained in section~\ref{zerosec}, the twist operator for a $\Z_2$ symmetry comes with zero modes. We can compute the charges carried by these zero modes by using the state-operator map and computing the charges of the corresponding state, following \eqn{table}. This has the advantage that we can exhibit the full $SU(2)_S\times U(1)_S\times U(1)_W\times U(1)_Y$ symmetries, rather than just the Cartan. Using \eqn{charge}, the lowest state $\kett{0}$ has charges
\be \kett{0}:  {\bf 1}_{0,1,-3}\ .\label{lonely}\ee
This is the charge of the operator that we call $T$ below. 
Then acting with a single zero mode we have the four states
\be q_L^{3-}\kett{0}: {\bf 1}_{-2,0,-2}\ \  , \ \ u_R^a\kett{0}:  {\bf 2}_{1,1,1}\ \  , \ \ l_L^-\kett{0}:  {\bf 1}_{0,0,-6}\ .\label{these}\ee
Acting with two zero modes gives six states, which are mutually conjugate,
%
\begin{equation}
\begin{alignedat}{6}
&q_L^{3-}u_R^a\kett{0}&&:\ &&  {\bf 2}_{-1,0,2}\ ,\quad && l_L^-u_R^a\kett{0}&&:\ &&{\bf 2}_{1,0,-2}\\
&q_L^{3-} l_L^-\kett{0}&&:\ && {\bf 1}_{-2,-1,-5}\ ,\quad &&  \epsilon_{ab}u_R^au_R^b\kett{0}&&:\ && {\bf 1}_{2,1,5} \ .
\end{alignedat}
\end{equation}
Acting with three zero modes gives the states conjugate to \eqn{these}
\be  \epsilon_{ab}u_R^au_R^bl_L^- \kett{0}: {\bf 1}_{2,0,2} \ \ , \ \  q_L^{3-} l_L^- u_R^a\kett{0}: {\bf 2}_{-1,-1,-1}\ \ ,   \ \ \epsilon_{ab}u_R^au_R^bq_L^{3-}\kett{0}: {\bf 1}_{0,0,6}  \ .\label{oddthisone}\ee
Finally, acting with all four zero modes gives the state conjugate to \eqn{lonely}
\be \epsilon_{ab}u_R^au_R^b q_L^{3-}l_L^-\kett{0}: {\bf 1}_{0,-1,3} \ .\ee
This completes the sixteen degenerate 2d ground states  in the presence of a  twist operator. Each of them has dimension 1/2 and so, from the perspective of two-dimensional scattering, acts like a free fermion.

\para
For our purposes, the outgoing scattering states are created by the twist operators \eqn{these} which have a single zero mode excited. This, then, is the result we're looking for: consistent scattering off a monopole is achieved by acting with the  twist operator $T$, that we take to have quantum numbers ${\bf 1}_{0,1,-3}$, as in \eqn{lonely}, together with an additional fermion zero mode. The paradoxical scattering \eqn{allwrong} then becomes
%
%
\begin{equation}\label{answer}
\begin{alignedat}{4}
&q_L^{a+} \ &&\longrightarrow\ u_R^a &&+ T\\
&d_R^3\ &&\longrightarrow\ q_L^{3-} &&+ T\\
&e_R\ &&\longrightarrow \ \ l_L^-\!&&+T\ .
\end{alignedat}
\end{equation}
Each of these scattering processes preserves the $SU(2)_S\times U(1)_S\times U(1)_W\times U(1)_Y$ symmetry.

\para
We can also compute the quantum numbers of these states under baryon and lepton number. In the realistic situation where $G$ is gauged, these famously suffer an ABJ anomaly, albeit with the amplitude so suppressed that no violation of these quantum numbers has been observed. If we alternatively view $G$ as a global symmetry, as we are here, then $B$ and $L$ have a mixed 't Hooft anomaly with $G$. This means that $B$ and $L$ are good quantum numbers in the bulk, but are not preserved by scattering off the monopole.

\para
We may easily determine the quantum numbers of the twist operator under $B$ and $L$ using the same technology that we reviewed previously. We find that we have 
\be B|0\rangle = L|0\rangle= -\frac{1}{2}|0\rangle\ .\ee
Correspondingly, the twist operator $T$ has $B=L=-\frac{1}{2}$. It is neutral under $B-L$.

\para
Curiously, the scattering processes above can be written in fully covariant fashion if we assign the twist operator $T$ quantum numbers $({\bf 1},{\bf 2})_{-3}$ under $SU(3)_S\times SU(2)_W\times U(1)_Y$ under the full Standard Model group. This is the same quantum numbers as the Higgs. It is not clear what it means to assign the twist operator such quantum numbers, given that the monopole breaks $G$. Nonetheless, it allows for the suggestive processes $q_L \longrightarrow u_R + T$ and $d_R\longrightarrow q_L +T$ and $e_R\longrightarrow l_L+T$.

\para
Scattering the anti-particles off the monopole reveals an interesting feature. We have
%
%
\begin{equation}\label{aanswer}
\begin{alignedat}{9}
&\bar{q}_L^{a+} \ &&\longrightarrow\ q_L^3 &&\ +\ && l_L^- &&\ +\ && \epsilon^{ab}u_R^b&&\ +\ && T \\
&\bar{d}_R^3\ &&\longrightarrow\ u_R^1&&\ +\ &&u_R^2&&\ +\ &&\ \ l_L^- &&\ +\ && T\\
&\bar{e}_R\ &&\longrightarrow \ u_R^1&&\ +\ &&u_R^2&&\ +\ &&\ \ q_L^{3-}&&\ +\ &&T\ .
\end{alignedat}
\end{equation}
This now looks like it's $1\ra 3$ scattering, as opposed to the $1\ra 1$ scattering seen in \eqn{answer}, but that's misleading. All the fermions on the right-hand side are zero modes of the twist operator and the outgoing state still has dimension 1/2 (viewed from the perspective of 2d scattering). It should be viewed as a single fermion. Again, both $B$ and $L$ are violated by a fractional amount, while $B-L$ is preserved. 

\para
Most generally, the scattering of $n_i$ ingoing fermions results in $\tilde{n}_i$ outgoing fermions, together with a twist operator
\be
  n_1 q_L^{1+} + n_2 q_L^{2+} + n_3 d_R^3 + n_4 e_R \ &\longrightarrow& \ \tilde{n}_1 q_L^{3-} + \tilde{n}_2 u_R^1 + \tilde{n}_3 u_R^2 + \tilde{n}_4 l_L^- + T^{n_1+n_2+n_3+n_4}\nn
\ee
where $\tilde{n}_i = \lceil (\tilde{Q}^{-1}Q)_{ij}n_j\rceil$ with $Q$ and $\tilde{Q}$ given in \eqn{qandtq}. More explicitly, 
\begin{align}
  \tilde{n}_1 &= \lceil \tfrac{1}{2}(-n_1-n_2+n_3-n_4) \rceil		\nn\\
  \tilde{n}_2 &= \lceil \tfrac{1}{2}(+n_1-n_2-n_3-n_4) \rceil		\nn\\
  \tilde{n}_3 &= \lceil \tfrac{1}{2}(-n_1+n_2-n_3-n_4) \rceil		\nn\\
  \tilde{n}_4 &= \lceil \tfrac{1}{2}(-n_1-n_2-n_3+n_4) \rceil		
\end{align}
and negative values for the integers $n_i$ or $\tilde{n}_i$ signify anti-particle rather than particle excitations. It is straightforward to check that $B-L$ is preserved.

\subsubsection*{Scattering Off the Anti-Monopole}

There is a similar story for scattering off the minimal anti-monopole \eqn{anticando}, with the ingoing and outgoing modes exchanged (up to a Weyl transformation)
%

\begin{equation}
\begin{array}{lc|cccc}
&&SU(2)_S&U(1)_S&U(1)_W& U(1)_Y\\\hline
\text{Incoming:}\quad&q^{1+}_L&\boldsymbol 1&2&1& 1\\
&u^a_R&\boldsymbol2 & -1 & 0 & 4\\
&l_L^+&{\boldsymbol 1}&0 & 1 & -3\\\hline
\text{Outgoing:}\quad& d_R^1&\boldsymbol1& 2 & 0 & -2\\
&q_L^{a-}&\boldsymbol2&-1 & -1 & 1\\
&e_R&{\boldsymbol1}&0 & 0 & -6
\end{array}
\end{equation}
now with $a=2,3$ the colour label. The story proceeds as before, just with one small difference. We need to determine the $\Z_2$ symmetry that acts on the outgoing fermions  $\psi_i = (q_L^{2-},q_L^{3-},d_R^1,e_R)$, leaving the ingoing fermions $\tilde{\psi}_i=(q_L^{1+},u_R^2,u_R^3,l_L^+)$ untouched. There are again two such embeddings of $\Z_2$ in $(U(1)_{B-L}\times G)/\Z_3$, and their generators are given by $\tilde{g}=(-1)^F w gw^{-1}$ and $\tilde{g}'=(-1)^F w g'w^{-1}$ with $g$ and $g'$ defined in \eqn{g1} and \eqn{g2} respectively, and $w$ an appropriate Weyl group transformation. 

\para
The story now proceeds as for the monopole. There is a 16-fold degeneracy of twist operators. We isolate one that we call $\tilde{T}$ that transforms as ${\bf 1}_{0,1,3}$. Those that we need for the monopole scattering are dressed with a single zero mode. The upshot is that we have the scattering
%
\begin{equation}
\begin{alignedat}{4}
&q_L^{1+}\ &&\longrightarrow\ d_R^1 &&+ \tilde{T}\\
&u_R^a \ &&\longrightarrow\ q_L^{a-} &&+ \tilde{T}\\ 
&l_L^+\! \ &&\longrightarrow\ e_R &&+ \tilde{T}\ .
\end{alignedat}
\end{equation}
Similar results hold for scattering multiple fermions and anti-fermions.

\subsubsection*{More Generations}

A natural (and phenomenologically relevant) generalization is to allow for $N_f$ generations of the particles above (where $N_f=3$ for our world). When doing so, there are new symmetries that need to be taken into account.

\para
For $N_f$ generations, in the absence of Yukawa coupling, there is an additional $SU(N_f)^5$ classical symmetry, under which each of the multiplets $l_L$, $q_L$, $u_R$, $d_R$ and $e_R$ transform independently. Some of these suffer a mixed anomaly with $U(1)_Y$ but there are different $SU(N_f)\times SU(N_f)$ subgroups that survive. Which, if any, is actually preserved by the monopole depends on the choice of UV completion.

\para
For example, there is a non-anomalous $SU(N_F)_B\times SU(N_f)_L$ under which the quarks $q_L$, $u_R$, and $d_R$ all transform in $({\bf N}_f,{\bf 1})$ while the leptons $l_L$ and $e_R$ transform in the $({\bf 1},{\bf N}_f)$. In this case, it's straightforward to see that the scattering \eqn{answer} and \eqn{aanswer} preserves both this non-Abelian flavour symmetry and the twist operator remains neutral. 

\para
Conversely, there is a different choice of non-anomalous $SU(N_f)_1\times SU(N_f)_2$ under which $q_L$, $u_R$ and $e_R$ transform in the representation  $({\bf N}_f,{\bf 1})$ while $l_L$ and $d_R$ transform in the $({\bf 1},{\bf N}_f)$. Indeed, this is the flavour symmetry group that arises in an $SU(5)$ GUT. It's more challenging to see how the scattering \eqn{answer} and \eqn{aanswer} is consistent with this symmetry. We leave this as an open question. 






\section{What Does it Mean?}\label{whatsec}

The resolution to the fermion-monopole scattering paradox proposed in \cite{vanBeest:2023dbu}  and, for the Standard Model, in \eqn{answer} and \eqn{aanswer} is that the outgoing fermions are created by a twist operator that is attached to a topological surface. But that's a rather abstract  and unfamiliar concept. In this section, we try to answer the question: what does it all mean? We will  also highlight some open problems, both interpretational and calculational.

\subsection{A 2d Toy Model}

In building up to this answer, we will first address the same question in two simple toy models. The first of these was the one that we leaned on heavily in \cite{vanBeest:2023dbu}: it is a two-dimensional theory of massless Weyl fermions. Two left-moving fermions carry charges 3 and 4 under a $U(1)$ symmetry while two right-moving fermions carry charges 0 and 5. The importance of these numbers lies in the fact that the $U(1)$ is anomaly free, courtesy of 
\be 3^2 + 4^2 = 0^2 + 5^2\ .\ee
Of course, free fermions have many different $U(1)$ symmetries, but we will imagine some irrelevant interactions that ensure that this particular chiral $U(1)$ is preserved.

\para
We can put this system on a spatial half-line, with a conformal boundary on the left that preserves the $U(1)$. Indeed, it is straightforward to construct the boundary state that achieves this \cite{Smith:2019jnh}. That leaves us with a puzzle that is very similar to the one that arises in fermion-monopole scattering: if you throw a charge 3 particle at the wall, then what bounces back? Naively, the only right-moving excitations carry charge 0 and 5 and it appears that nothing can bounce back. But one can show that what actually bounces back is two fermions attached to twist operator $T$ that enacts a $\Z_5$ symmetry \cite{vanBeest:2023dbu}

\begin{equation}
\begin{tikzpicture}[baseline=.9cm]
 
\draw[very thick] (-3,0) -- (-3,2);
\draw[->,>=stealth,thick] (-1.2, 0.2) -- (-2.2, 0.7);
\draw[<-,>=stealth,thick] (-1.2, 1.8) -- (-2.2, 1.3);
\draw[decorate, decoration={snake,amplitude=.6mm}] (-3,1) -- (-2.2, 1.3);
 
\foreach \x in {0,...,12} \draw (-3-.2,.15*\x+.04) -- (.2-3-.2,.15*\x+.1+.04);
\foreach \x in {-1,...,11} \draw (-3-.1,.15*\x+.09+.075) -- (.2-3-.2,.15*\x+.1+.04+.075);
 
\node at (-.1, 1.9) {$\tilde{\psi}_0 +\tilde{\psi}_5 + T$};
\node at (-1.7+1,.1) {$\psi_3$};
 
\filldraw (-1.2, 0.2) circle (1pt);
\filldraw (-2.2, 1.3) circle (1pt);

\end{tikzpicture}
\end{equation}
This twist operator $T$ carries charge $-2$ under the $U(1)$, ensuring that charge is conserved.

\para
Imagine that you are an experimenter in this world. You sit far from the wall and can perform any experiment you choose. So, for example, you can compute the heat capacity and convince yourself that you live in a world with $c_L=c_R=2$. You can create sets of particles and anti-particles (consistent with the $U(1)$ symmetry) and measure their charge by seeing how they are accelerated through regions of non-vanishing chemical potential. By performing such experiments, you learn that left-moving fermions carry charges 3 and 4, and right-moving fermions carry charges 0 and 5. 

\para
Then, one day, you perform a novel experiment: you throw a charge 3 particle to the left and you wait. At some point, to your surprise, a particle returns. This is the combination $\tilde{\psi}_0+\tilde{\psi}_5+T$. What do you measure?

\para
As written, this looks like a two fermion state (plus the twist operator). But that's  misleading. As shown in \cite{vanBeest:2023dbu}, the combination of outgoing modes $\tilde{\psi}_0+\tilde{\psi}_5+T$ is an operator with dimension 1/2. In other words, it doesn't look like a composite at all, but like a free fermion. But, crucially, it's a right-moving free fermion with charge +3.  The  fact that we're in $d=1+1$ dimensions helps with this interpretation because the ``components'' $\tilde{\psi}_0$, $\tilde{\psi}_5$ and $T(x)$ all move together to the right at the speed of light. This is in contrast to  higher dimensions where it's much easier for massless particles to separate simply by moving in different directions.

\para
This should come as a surprise. After many decades of painstaking experiments, you have convinced yourself that right-moving fermions carry charges 0 and 5 and your measurements of heat capacity tell you that there is no extra right-moving fields that you missed. Yet here is a right-moving charge 3 fermion. 

\para
You might wonder if there is something strange about this fermion that you can detect. Your theorist colleagues mumble words about ``twist operators'' and tell you that, in Euclidean space, these operators have the property that they are attached to a topological line so that correlation functions pick up a phase as operators are adiabatically moved through this line. None of which sounds very measurable! Indeed, in Lorentzian signature the topological line divides spacetime in two. That makes it very hard to see how one can do an Aharonov-Bohm type experiment to detect it in this gapless system. (Such an experiment might be possible if we keep the system gapless, but break Lorentz invariance so that the trailing branch cut is no longer dragged along the lightcone.)

\para
The conclusion in this toy model is rather stark: by bouncing  a charge 3 fermion off the edge of the universe, it can travel the ``wrong way'' even though there is no right-moving field in the (untwisted) Fock space of which it is an excitation.

\para
Suppose that we repeat this experiment five times, with a suitable delay between them. We now have five right-moving particles, each carrying charge 3. Now there is no longer a topological defect line stretching to the boundary and, indeed, viewed from afar the total charge is $5\times 3$ which is  a perfectly good excitation in the original Hilbert space.  But it's an excitation that is created by local operators with topological defect lines stretched between them, instead of stretched to the boundary. It may well be that a suitably ingenious local experimental apparatus could manufacture such a state, although we don't know how. The boundary gives a simple way to do so. 

\para
This story also seems to suggest a dual description where the right-moving charge 3 fermion is a fundamental field and the charge 0 and 5 fields are twist operators. At the end of the day, twist fields are locally indistinguishable from regular fermions so it makes sense that such a dual description exists. The fact that the twist field has non-trivial holonomy around an untwisted field means that they cannot be both local in a single description, but it doesn't say which is twisted and which untwisted. For example, in the Ising model the spin operator $\sigma$ and disorder operator $\mu$ are interchanged under duality~\cite{PhysRev.60.252}. 

\para
One further complication is that the scattering takes place in a gapless theory and it is difficult to get intuition for what it means to localise and/or scatter charged gapless degree of freedom. It may well be that a better handle is needed on the IR divergences of the theory \cite{Kinoshita:1962ur,Lee:1964is}. One interesting avenue would be to see if the properties of the twisted sector survive under symmetric mass generation of the kind proposed in \cite{Wang:2013yta,Wang:2018ugf,Tong:2021phe}. 

\para
Before we move on to 4d, it's worth commenting that there are experiments that share some, but not all, of the aspects of this problem. The two-channel Kondo effect also exhibits scattering into a twisted sector \cite{Maldacena:1995pq} and was realised experimentally in \cite{Potok_2007}. In the context of $d=2+1$ dimensions, the scattering of anyons off a barrier to create an anyon-electron pair, as seen in \cite{Glidic_2023}, provides an example of twisted states (the anyons) scattering into non-twisted states (the electron). 

\subsection{Scattering in 4d}

Now we can ask the same questions for our simple 4d model \eqn{sqed}  with two left-handed fermions transforming as
\begin{equation}
\begin{array}{l|cc}
&U(1) &SU(N)\\\hline
\psi\quad& +1 & {\bf N} \\ 
\tilde{\psi} & -1 & {\bf N} \\
\end{array}\label{oldone}
\end{equation}
%
%
%
In the presence of a monopole, an ingoing fermion $\psi$ with charge ${\bf N}_{+1}$ scatters into an outgoing fermion $\tilde{\psi}$ with charge ${\bf N}_{-1}$, preserving $U(1)\times SU(N)$. As we have seen in \eqn{twist}, the outgoing fermion is attached to a twist operator $T^{-2}$ with charge $+2$.

\para
Viewed from the two-dimensional perspective, the story is the same as that told above: the combination of $\tilde{\psi} + T^{-2}$ has dimension 1/2, and so acts just like a free fermion moving in the ``wrong'' direction. From the 4d perspective, this translates to the fact that we seemingly have a fermion of the ``wrong'' handedness. The outgoing fermion has the properties of a \emph{right-handed} fermion transforming as ${\bf N}_{+1}$. This is not part of the original Fock space. 

\para
This suggests that scattering off the monopole and anti-monopole gives a parity-doubling of the spectrum of the theory! There are, however, some issues to address.

\para
First, the outgoing particle with the ``wrong'' quantum numbers sits in the s-wave. But physics is local and a particle detector does not measure particles in an angular momentum eigenstate. Instead, it registers a definite trajectory. Can one form a localised particle in the twisted sector?

\para
The answer appears to be yes, although the details are somewhat confusing.  The $\Z_N$ symmetry that implements the twist operator acts on all higher angular momentum modes, not just the s-wave. But, for these, $\Z_N$ acts on both ingoing and outgoing modes, shifting the quantum numbers of both. This suggests  that the theory has the full complement of angular momentum modes in the twisted sector, even though only the twisted s-wave is created in monopole scattering. Presumably the right way to think about a localised particle is as a superposition of many topological surface operators, in the same way that the localised particle is a superposition of many modes on the sphere. Clearly it would be good to understand this issue better.

\para
Scattering in the Standard Model brings one further level of confusion. The 
$\Z_2$ symmetry \eqn{g1} for the monopole sits entirely within the gauge group $SU(3)\times SU(2)\times U(1)_Y$. (The alternative $\Z_2$ symmetry defined in \eqn{g2} is gauge equivalent to $(-1)^F$ so fares slightly better, but we could always consider scattering in the Standard Model in which $B-L$ is gauged and the problem rears its head once more.) Certainly if the defect operator acted on a closed manifold then this would be cause for concern. In our case, the operator acts on an open ball  and the result is something akin to the Gukov-Witten surface operator \cite{Gukov:2006jk}, which is a disorder operator defined by cutting out a codimension-2 submanifold along which the gauge field is singular, and specifying a boundary condition for the gauge field as we approach the singular locus. The reduction to 2d shows that, in the present context, the boundary of the ball carries electric charge.

\para
There is another, related, way in which the story for the Standard Model differs from the theory \eqn{oldone} that we discussed in \cite{vanBeest:2023dbu}. In the previous paper, when the monopole had suitably large magnetic charge $g\geq 3$, the twist symmetry was non-invertible in $d=3+1$ dimensions and the associated defect operator coincided with the ABJ defect introduced in \cite{Choi:2022jqy,Cordova:2022ieu}. In particular, this means that the defect was necessarily accompanied by a 3d TQFT. For low charge monopoles, $g=1$ and $g=2$, the natural generalisation was to stack the defect with an SPT phase (essentially an integer quantum Hall state). In both cases, this offered an independent way to compute the charge of the defect by considering Wilson lines in the associated Chern-Simons theory that lives on the defect. A full understanding of the analogous story for the Standard Model is still lacking.

\subsection{More Open Questions}

A recurring theme in this (and previous) work is that, whenever a theory contains chiral excitations that carry different quantum numbers under various symmetries, scattering processes become subtle, since it is not always obvious how to write down a candidate out-state in such a way that all conservation laws are obeyed. The general mechanism we have described here is that the out-state is typically dressed by a surface in one higher dimension, which adds charge under the symmetries, allowing us to find an excitation with the right quantum numbers. In this regard, there is an important distinction between abelian and non-abelian symmetries. The former are well understood, with twist operators contributing to the charge as shown in \cite{vanBeest:2023dbu} and reviewed earlier in this paper. In contrast, non-abelian symmetries are trickier: for one thing, the twist vacuum of an invertible topological line that commutes with a semi-simple group is \emph{always neutral under such a group}, as long as it does not carry zero-modes. This is because if the twist vacuum is unique, it must transform as a one-dimensional representation of the semi-simple group, but all such representations are trivial (see also footnote 19 in~\cite{vanBeest:2023dbu}). One of the take-away messages from the analysis of section~\ref{monosec} is that zero-modes generically do endow the twist vacua with a charge under non-abelian symmetries.

\para
In absence of zero-modes, then, topological lines can only add abelian charge. Therefore, if the in- and out-modes carry different non-abelian charges, the addition of a branch cut will not solve the mismatch, unless the topological line yields zero-modes for some of the fermions.

\para
A more subtle phenomenon is the fact that ingoing and outgoing waves usually not only carry different representations under non-abelian groups, but also these representations have different N-ality. For example, in the 1-5-7-8-9 model described in section~\ref{examplesec}, there are two ingoing modes (with charges $7$ and $8$) and three outgoing modes (with charges $1,5,9$). The charge-$8$ fermion is charged under $\mathbb Z_2=Z(SU(2)_\text{rot})$ (as it transforms under an even-dimensional representation) but the outgoing modes are all neutral (being odd-dimensional). This mismatch cannot be fixed by adding a topological line, \emph{even if this line yields fermion zero-modes}. Indeed, if a fermion transforms under the representation $R$ under some semi-simple group $G$, then the zero-modes transform as $\wedge^n R$, for $n=0,1,\dots,\dim R$. It is clear that, if $R$ does not transform under some $\Gamma\subseteq Z(G)$, then neither does $\wedge^n R$.\footnote{A more abstract argument is: if the zero-modes carry more N-ality than the fermions themselves, then the $G$ symmetry would be realised projectively, i.e., the effective $0+1$ system would have an 't Hooft anomaly. But this is not possible for complex fermions since we can always write down a $G$-symmetric mass term. By contrast, real fermions can and typically do give rise to zero-modes with higher N-ality (and, relatedly, the Hilbert space of real fermion zero-modes is no longer the representation space of $\{\wedge^n R\}$, see appendix B in~\cite{Delmastro:2022pfo} for more details). In this work we always begin with 4d fermions, which are complex, and perform s-wave reduction, which always yields complex 2d fermions. It would be interesting to understand the various charges of fermion zero-modes in some context where dimensional reduction yields real 2d fermions.\label{fn:anomaly}} This class of chiral models, then, leads to more exotic out-states: a twist of the same type we have been considering so far does not work.

\para
What other options are available for us to construct a valid out-state? Here we list some possibilities:\footnote{We would like to thank Zohar Komargodski and Shu-Heng Shao for helpful discussions regarding some of the points below.}
\begin{itemize}
\item First, the twist may be non-invertible even after s-wave reduction. This is certainly a valid option in 2d: generic boundary conditions map local operators into twist operators, and the corresponding topological line can definitely be non-invertible. That being said, it is not clear whether this is a valid option in 4d: generically, non-invertible zero-form symmetries in $d>2$ always act invertibly on local operators, except possibly when the theory has a $(d-2)$-form symmetry. Therefore, unless generic chiral gauge theories have suitable 2-form symmetries, this doesn't appear to be a valid option.


\item Second, the out-state could be twisted by some defect that does not commute with $G$. This option is unlikely to lead to a convincing picture, since the endpoint of a line carries well-defined $G$-charge only when the line commutes with $G$. Otherwise, the twisted Hilbert space does not realise a representation of $G$ -- the twisted boundary condition explicitly breaks the symmetry. When $G=SU(2)_\text{rot}$, this would mean that the out-going states do not have well-defined Lorentz spin, which sounds unnatural given that every other ingredient in this story is spherically symmetric (and it might even invalidate s-wave reduction).

\item Finally, it might simply be the case that no $G$-symmetric (simple) boundary condition exists, even if $G$ is free of 't Hooft anomalies. The 4d interpretation of this situation is that there is no $G$-symmetric (simple) monopole at all, and any UV completion where the monopole is dynamical must either break $G$ at some scale \emph{or} monopoles preserving $G$ at all scales must have local degrees of freedom associated with them in the IR.

\end{itemize}
\noindent We point out that this discussion is not merely academic: some monopoles in the Standard Model give rise to a spectrum of fermions whose charges under $SU(2)_\text{rot}$ are precisely as above. For example, the natural charge-3 monopole $(n_1,n_2,m,g)=(0,0,1,0)$, leads to the following scattering states:
\begin{equation}
\begin{array}{lc|cccc}
&&SU(3)&U(1)_W&U(1)_Y&SU(2)_\text{rot}\\\hline
\text{Incoming:}\quad&q_L&\boldsymbol3&1&1&\boldsymbol1\\
&\bar e_R&\boldsymbol1&0&6&\boldsymbol3\\
&\bar d_R&\overline{\boldsymbol3}&0&2&\boldsymbol1\\\hline
\text{Outgoing:}\quad&l^+_L&\boldsymbol1&1&-3&\boldsymbol1\\
&l^-_L&\boldsymbol1&-1&-3&\boldsymbol2\\
&\bar u_R&\overline{\boldsymbol3}&0&-4&\boldsymbol2
\end{array}
\end{equation}
As advertised, one chirality contains only odd-dimensional $SU(2)_\text{rot}$ representations, while the other chirality has even-dimensional representations as well. Scattering these leads to the aforementioned exotic states that we do not know how to describe. One might be tempted to simply declare that this monopole is unphysical, but this is hardly a valid excuse. Indeed, this charge-3 monopole is in fact one of the stable, spherically symmetric monopoles in the usual $SU(5)$ GUT model~\cite{PhysRevLett.52.879}, and therefore a perfectly healthy UV completion does exist. It seems that there is much more still to learn about monopoles in theories of chiral fermions.

\appendix

\section{Analysis of Stability}\label{stability}

Scalar fields in the vicinity of a 't Hooft line experience a centrifugal barrier. In the case of fermions, the lowest lying angular momentum mode does not encounter such a barrier, as we have seen. Spin-1 fields are yet stranger. Under certain circumstances, a charged spin-1 field experiences an \textit{attractive} potential near a 't Hooft line. Such fields---which arise as off-diagonal W-bosons in any non-Abelian theory---are thus sucked into the 't Hooft line, signalling an instability \cite{Aharony:2022ntz,Aharony:2023amq,Shnir:2005vvi}. For each value of the topological charge $p\in \Z$, we expect only a finite number of stable monopoles $(n_1,n_2,m,g)$, with $p=2n_1+2n_2+3m+6g$.

\para
In detail, suppose a theory has a spin-1 field which experiences a charge times magnetic flux $c\in \Z$; that is, it is a section of a $U(1)$ vector bundle of first Chern class $c$. Then, for all $|c|\ge 2$, the lowest lying angular momentum mode experiences an attractive potential, and the monopole is unstable. Only the minimal monopoles $c=\pm 1$ (as well as the trivial case $c=0$) are stable.

\para
This kind of stability analysis is very sensitive to the global form of the gauge group. For example, consider a 't Hooft line in a theory with gauge algebra $\frak{su}(2)$. The corresponding field strength is
\be
  F = \begin{pmatrix}
  	\frac{1}{2}m		&		\\
  	&	-\frac{1}{2}m
  \end{pmatrix} \left(\frac{1}{2} \sin \theta \,d \theta \wedge d \phi\right)
\nn\ee
for some $m\in \Z$, where $m\ge 0$ without loss of generality. If the gauge group is $SO(3)$, then we can have any $m\in\Z$. If it's $SU(2)$, then only $m\in 2\Z$ is allowed. The off-diagonal W-bosons have $c=\pm m$. We see that in the $SU(2)$ theory there are no stable monopoles, which is consistent with the trivial magnetic 1-form symmetry. Conversely, the $SO(3)$ theory has a single stable monopole, with $m=1$, which carries charge under the $\Z_2$ magnetic 1-form symmetry.

\para
We return to the $(n_1,n_2,m,g)$ monopoles of the Standard Model. There are then four charge-conjugate pairs of charged W-bosons to worry about. Those from the off-diagonal of $SU(2)$ have $c=\pm m$. Meanwhile, those from the off-diagonal of $SU(3)$ have $c=\pm n_1,\pm n_2,\pm(n_1-n_2)$. It follows that stability requires
\begin{align}
	n_1,n_2,m\le 1
\end{align}
while $g$ can be arbitrary. We thus learn that for each value of the 1-form magnetic charge $p=2n_1+2n_2+3m+6g$ there is precisely \textit{one} stable monopole. The non-Abelian magnetic charges of this stable monopole are
\begin{align}
	(n_1,n_2,m) = \left\{\begin{aligned}
\,\,(0,0,0)\qquad 		& p = 0 \,\, (\text{mod } 6) \\
\,\,(1,1,1)\qquad 				& p = 1 \,\, (\text{mod } 6) \\
\,\,(1,0,0)\qquad 				& p = 2 \,\, (\text{mod } 6) \\
\,\,(0,0,1)\qquad 				& p = 3 \,\, (\text{mod } 6) \\
\,\,(1,1,0)\qquad 				& p = 4 \,\, (\text{mod } 6) \\
\,\,(1,0,1)\qquad 				& p = 5 \,\, (\text{mod } 6) 
\end{aligned}
 \right.
\end{align}

\section{Appendix: More solvable monopoles}\label{solvesec}

In the main text, we considered the scattering of fermions off the minimal monopole, as well as the corresponding anti-monopole. These have topological class  $\pm 1$, and the scattering off each reduces to a two-dimensional boundary CFT problem with $c_L=c_R=4$. What made these monopoles solvable was the fact that we had enough symmetry to entirely constrain the problem.

\para
We  can solve the scattering problem whenever the number of fermions in the lowest angular momentum modes is equal to the rank of the preserved symmetry group which, generally, is given by some rank 4 subgroup of $G$, together with $U(1)_{B-L}$ and $SU(2)_\text{rot}$ rotations. This means that we can solve the scattering whenever the number of fermions in lowest angular momentum modes is $c_L=c_R\leq 6$.  There then turns out to be precisely nine monopoles with this property, whose scattering we have a chance to solve completely\footnote{A priori, the constraints imposed by symmetry conservation could be degenerate, and thus not entirely solve the scattering problem. Happily, this turns out not to happen.}.

\para
We have already met two of these. There are two more with $c_L=c_R=4$, which have $(n_1,n_2,m,g)=(0,0,2,-1)$ and $(2,1,0,-1)$, and which both lie in the trivial topological class. Their scattering is solved in a qualitatively similar way to the monopoles of the main text; one finds that scattering states may be attached to a $\Z_2$ topological line, corresponding to a twist vacuum with $16$-fold degeneracy. We omit further details here.

\para
The remaining five monopoles have $c_L=c_R=6$. They have $(n_1,n_2,m,g)=(1,0,0,0)$ and $(1,1,0,-1)$ in class $\pm 2$, $(n_1,n_2,m,g)=(2,0,1,-1)$ and $(2,2,1,-2)$ in class $\pm 1$, and $(n_1,n_2,m,g)=(2,1,2,-2)$ in the trivial class. The scattering problem is solved in a similar way for each of these monopoles, albeit with some features that distinguish them from the $c_L=c_R=4$ cases. First, for each monopole, both the ingoing and outgoing sectors are made up of an $SU(2)_\text{rot}$ doublet, along with four singlets; in particular, there are modes charged under the centre of $SU(2)_\text{rot}$ in both ingoing and outgoing sectors. Moreover, the scattered states are generically connected to a topological line generating a $\Z_3$ symmetry. This means, among other things, that the corresponding twist vacua are unique.

\para
Here we present the full scattering solution for the $(n_1,n_2,m,g)=(2,0,1,-1)$ monopole; the rest are solved in a qualitatively similar fashion. This monopole sits in the same topological class $+1$ as the `minimal' monopole described in the main text, which means that it is possible to deform one into the other by continuously changing the magnetic field. Configurations along this path will necessarily break spherical symmetry, and potentially also $SU(2)_S$. 

\para 
The symmetries preserved by the monopole (neglecting discrete quotients) are
\begin{align}
   SU(2)_S \times U(1)_S \times U(1)_W \times U(1)_Y \times U(1)_{B-L}\times SU(2)_\text{rot}
\end{align}
The resulting two-dimensional theory has the following modes,
\begin{equation}
\begin{array}{lc|cccccc}
&&SU(2)_S&U(1)_S&U(1)_W& U(1)_Y & U(1)_{B-L} & SU(2)_\text{rot}\\\hline
\text{Incoming:}\quad& (q^{a+}_L)^A&\boldsymbol 2 & -2 & 1 & 1 & 1 & {\bf 1}\\
& q_L^{1-} & {\bf 1} & -2 & -1 & 1 & 1 & {\bf 1} \\
&d^a_R& {\bf 1} & 1 & 0 & -2 & 1 & {\bf 2} \\
&e_R&{\boldsymbol 1}&0 & 0 & -6 & -3 & {\bf 1} \\\hline
\text{Outgoing:}\quad& (u_R^1)^A &\boldsymbol2& -2 & 0 & 4 & 1 &{\bf 1}\\
& d_R^1& {\bf 1} & -2 & 0 & -2 & 1 & {\bf 1} \\
& q_L^{a-} & {\bf 1} & 1 & -1 & 1 & 1 & {\bf 2} \\
&l_L^-& {\bf 1} & 0 & -1 & -3 & -3 & {\bf 1}
\end{array}
\end{equation}
%
%
where $A=\,\uparrow,\downarrow$ are indices for the fundamental of $SU(2)_\text{rot}$. 
The  general scattering solution is then
\begin{align}
&n_1 (q_L^{1+})^\uparrow + n_2 (q_L^{1+})^\downarrow + n_3 q_L^{1-} + n_4 d_R^2 + n_5 d_R^3 +n_6 e_R  \\[1em]
& \longrightarrow \quad \tilde{n}_1 (u_R^{1})^\uparrow + \tilde{n}_2 (u_R^{1})^\downarrow + \tilde{n}_3 d_R^1 + \tilde{n}_4 q_L^{2-} + \tilde{n}_5 q_L^{3-}  + \tilde{n}_6 l_L^-+ T^{n_1+n_2-n_3+n_4+n_5+n_6}\nn
\end{align}
where $\tilde{n}_\alpha = \lfloor \frac{1}{2} + n_\alpha - \frac{1}{3}(n_1+n_2-n_3+n_4+n_5+n_6) \rfloor $ except for $\tilde{n}_3$, which is $\tilde{n}_3 = \lfloor \frac{1}{2} + n_3 + \frac{1}{3}(n_1+n_2-n_3+n_4+n_5+n_6) \rfloor $.

\para
Here $T$ is the twist operator living at end the of a topological line enacting a $\Z_3$ transformation, which acts only on outgoing modes as
\begin{align}
  q_L^{a-}	\,\,\longrightarrow\,\, e^{-2\pi i/3} q_L^{a-}	,\quad d_R^1 \,\,\longrightarrow\,\, e^{2\pi i/3} d_R^1,\quad u_R^{1A} \,\,\longrightarrow\,\, e^{-2\pi i/3} u_R^{1A},\quad l_L^-\,\, \longrightarrow\,\, e^{-2\pi i/3}l_L^-
\nn\end{align}
The corresponding twist operators $T, T^2$ sit in the representations $(\1,0,1,-3,0,\1)$ and $(\1,0,-1,3,0,\1)$, respectively. One can verify that scaling dimension is indeed conserved by this scattering.

\subsection*{Acknowledgements}

We are especially grateful to Zohar Komargodski for collaboration at an early stage of this project, for many discussions that helped orient our thinking, and for extensive comments on an early draft of the paper. Thanks also to Sumit Das and Cesar Gomez for useful discussions. 
This work was supported by the STFC grant ST/L000385/1, the EPSRC grant EP/V047655/1 ``Chiral Gauge Theories: From Strong Coupling to the Standard Model'', and a Simons Investigator Award. PBS is supported by WPI Initiative, MEXT, Japan at Kavli IPMU, the University of Tokyo.

\printbibliography

\end{document}